\documentclass{aastex701}

\usepackage{graphicx}
\usepackage{subfig}
\usepackage{booktabs}
\usepackage{amsmath}
\usepackage{appendix}

\begin{document}

\title{Evidence for Enhancement in the Rate of Fast Radio Bursts Toward Galaxy Clusters}

\author[orcid=0000-0002-4623-5329,sname=Sammons, gname=Mawson]{Mawson W. Sammons}
\affiliation{Department of Physics, McGill University, 3600 rue University, Montr\'eal, QC H3A 2T8, Canada}
\affiliation{Trottier Space Institute, McGill University, 3550 rue University, Montr\'eal, QC H3A 2A7, Canada}
\affiliation{Laboratoire d'Astrophysique de Marseille, Aix-Marseille Univ., CNRS, CNES, Marseille, France}
\email[show]{mawson.sammons@mcgill.ca} 

\author[0009-0006-8812-9109]{Airene Ahuja}
\affiliation{Trottier Space Institute at McGill University, 3550 rue University, Montr\'eal, QC H3A 2A7, Canada}
\affiliation{Department of Physics, McGill University, 3600 rue University, Montr\'eal, QC H3A 2T8, Canada}
\email{aahuja07@student.ubc.ca}

\author[orcid=0000-0001-7166-6422, sname=Dobbs, gname=Matt]{Matt Dobbs} 
\affiliation{Department of Physics, McGill University, 3600 rue University, Montr\'eal, QC H3A 2T8, Canada}
\affiliation{Trottier Space Institute, McGill University, 3550 rue University, Montr\'eal, QC H3A 2A7, Canada}
\email{matt.dobbs@mcgill.ca}

\author[orcid=0000-0003-2739-5869, sname=Kader, gname=Zarif]{Zarif Kader}
\affiliation{Department of Physics, McGill University, 3600 rue University, Montr\'eal, QC H3A 2T8, Canada}
\affiliation{Trottier Space Institute, McGill University, 3550 rue University, Montr\'eal, QC H3A 2A7, Canada}
\email{zarif.kader@mail.mcgill.ca}

\author[orcid=0009-0002-1944-6398, sname=Davies-Velie, gname=Evan]{Evan Davies-Velie} 
\affiliation{Department of Physics, McGill University, 3600 rue University, Montr\'eal, QC H3A 2T8, Canada}
\affiliation{Trottier Space Institute, McGill University, 3550 rue University, Montr\'eal, QC H3A 2A7, Canada}
\email{evan.davies-velie@mail.mcgill.ca}


\author[0000-0002-3615-3514]{Mohit Bhardwaj}
\affiliation{Department of Physics, Indian Institute of Technology Kanpur, Kalyanpur, Kanpur, Uttar Pradesh 208016, India}
\email{mohitb@iitk.ac.in}

\author[0000-0002-1800-8233]{Charanjot Brar}
\affiliation{National Research Council of Canada, Herzberg Astronomy and Astrophysics, 5071 West Saanich Road, Victoria, BC V9E 2E7, Canada}
\email{charanjot.brar@nrc-cnrc.gc.ca}

\author[0000-0001-6422-8125]{Amanda M. Cook}
\affiliation{Department of Physics, McGill University, 3600 rue University, Montr\'eal, QC H3A 2T8, Canada}
\affiliation{Trottier Space Institute at McGill University, 3550 rue University, Montr\'eal, QC H3A 2A7, Canada}
\affiliation{Anton Pannekoek Institute for Astronomy, University of Amsterdam, Science Park 904, 1098 XH, Amsterdam, The Netherlands}
\email{amanda.cook@mail.mcgill.ca}

\author[0000-0002-8376-1563]{Alice P.~Curtin}
  \affiliation{Department of Physics, McGill University, 3600 rue University, Montr\'eal, QC H3A 2T8, Canada}
  \affiliation{Trottier Space Institute, McGill University, 3550 rue University, Montr\'eal, QC H3A 2A7, Canada}
  \affiliation{Anton Pannekoek Institute for Astronomy, University of Amsterdam, Science Park 904, 1098 XH Amsterdam, The Netherlands}
  \email{alice.curtin@mcgill.ca}

\author[0000-0002-3382-9558]{B.~M.~Gaensler}
  \affiliation{Department of Astronomy and Astrophysics, University of California, Santa Cruz, 1156 High Street, Santa Cruz, CA 95060, USA}
  \affiliation{Dunlap Institute for Astronomy \& Astrophysics, University of Toronto, 50 St.~George Street, Toronto, ON M5S 3H4, Canada}
  \affiliation{David A.~Dunlap Department of Astronomy \& Astrophysics, University of Toronto, 50 St.~George Street, Toronto, ON M5S 3H4, Canada}
  \email{gaensler@ucsc.edu}

\author[0009-0001-2196-8251]{Affan Khadir}
\affiliation{Trottier Space Institute at McGill University, 3550 rue University, Montr\'eal, QC H3A 2A7, Canada}
\affiliation{Department of Physics, McGill University, 3600 rue University, Montr\'eal, QC H3A 2T8, Canada}
\email{affan.khadir@mail.mcgill.ca}

\author[0000-0001-9345-0307]{Victoria M. Kaspi}
\affiliation{Trottier Space Institute at McGill University, 3550 rue University, Montr\'eal, QC H3A 2A7, Canada}
\affiliation{Department of Physics, McGill University, 3600 rue University, Montr\'eal, QC H3A 2T8, Canada}
\email{victoria.kaspi@mcgill.ca}

\author[0009-0004-4176-0062]{Afrokk Khan}
\affiliation{Trottier Space Institute at McGill University, 3550 rue University, Montr\'eal, QC H3A 2A7, Canada}
\affiliation{Department of Physics, McGill University, 3600 rue University, Montr\'eal, QC H3A 2T8, Canada}
\email{afrasiyab.khan@mcgill.ca}
  
\author[0000-0003-2116-3573]{Adam Lanman}
  \affiliation{MIT Kavli Institute for Astrophysics and Space Research, Massachusetts Institute of Technology, 77 Massachusetts Ave, Cambridge, MA 02139, USA}
  \affiliation{Department of Physics, Massachusetts Institute of Technology, 77 Massachusetts Ave, Cambridge, MA 02139, USA}
  \email{alanman@mit.edu}
  
\author[0000-0002-4209-7408]{Calvin Leung}
  \affiliation{Miller Institute for Basic Research, University of California, Berkeley, CA 94720, United States}
  \affiliation{Department of Astronomy, University of California, Berkeley, CA 94720, United States}
\email{}
  
\author[0000-0003-4584-8841]{Lluis Mas-Ribas}
  \affiliation{Department of Astronomy and Astrophysics, University of California, Santa Cruz, 1156 High Street, Santa Cruz, CA 95060, USA}
\email{lmr@ucsc.edu}
  
\author[0000-0002-4279-6946]{Kiyoshi W.~Masui}
  \affiliation{MIT Kavli Institute for Astrophysics and Space Research, Massachusetts Institute of Technology, 77 Massachusetts Ave, Cambridge, MA 02139, USA}
  \affiliation{Department of Physics, Massachusetts Institute of Technology, 77 Massachusetts Ave, Cambridge, MA 02139, USA}
\email{kmasui@mit.edu}

\author[0000-0003-2111-3437]{Kyle McGregor}
  \affiliation{Trottier Space Institute at McGill University, 3550 rue University, Montr\'eal, QC H3A 2A7, Canada}
  \affiliation{Department of Physics, McGill University, 3600 rue University, Montr\'eal, QC H3A 2T8, Canada}
\email{kyle.mcgregor@mail.mcgill.ca}
  
\author[0000-0002-2551-7554]{Daniele Michilli}
  \affiliation{Laboratoire d'Astrophysique de Marseille, Aix-Marseille Univ., CNRS, CNES, Marseille, France}
\email{danielemichilli@gmail.com}

\author[0000-0002-8897-1973]{Ayush Pandhi}
\affiliation{Trottier Space Institute at McGill University, 3550 rue University, Montr\'eal, QC H3A 2A7, Canada}
\affiliation{Department of Physics, McGill University, 3600 rue University, Montr\'eal, QC H3A 2T8, Canada}
\email{ayush.pandhi@mcgill.ca}
  
\author[0000-0002-8912-0732]{Aaron B.~Pearlman}
  \altaffiliation{NASA Hubble Fellow.}
  \affiliation{MIT Kavli Institute for Astrophysics and Space Research, Massachusetts Institute of Technology, 77 Massachusetts Ave, Cambridge, MA 02139, USA}
  \affiliation{Department of Physics, McGill University, 3600 rue University, Montr\'eal, QC H3A 2T8, Canada}
  \affiliation{Trottier Space Institute, McGill University, 3550 rue University, Montr\'eal, QC H3A 2A7, Canada}
\email{aaron.b.pearlman@mit.edu}

\author[0000-0002-4795-697X]{Ziggy Pleunis}
\affiliation{ASTRON, Netherlands Institute for Radio Astronomy, Oude Hoogeveensedijk 4, 7991 PD Dwingeloo, The Netherlands}
\affiliation{Anton Pannekoek Institute for Astronomy, University of Amsterdam, Science Park 904, 1098 XH, Amsterdam, The Netherlands}
\email{z.pleunis@uva.nl}

\author[0009-0008-2000-6959]{Sachin Pradeep E.~T.}
\affiliation{Trottier Space Institute at McGill University, 3550 rue University, Montr\'eal, QC H3A 2A7, Canada}
\affiliation{Department of Physics, McGill University, 3600 rue University, Montr\'eal, QC H3A 2T8, Canada}
\email{sachin.pradeepetakkepravanthulicheri@mail.mcgill.ca}

\author[0000-0002-8823-8835]{Aniket Prasad}
  \affiliation{Laboratoire d'Astrophysique de Marseille, Aix-Marseille Univ., CNRS, CNES, Marseille, France}
\email{aniket.prasad@lam.fr}

\author[0000-0002-7738-6875]{J.~Xavier Prochaska}
  \affiliation{Department of Astronomy and Astrophysics, University of California, Santa Cruz, 1156 High Street, Santa Cruz, CA 95060, USA}
\email{xavier@ucolick.org}

\author[0000-0002-7374-7119]{Paul Scholz}
\affiliation{Department of Physics and Astronomy, York University, 4700 Keele Street, Toronto, Ontario, ON MJ3 1P3, Canada}
\email{pscholz@yorku.ca}

\author[0000-0002-6823-2073]{Kaitlyn Shin}
\affiliation{Cahill Center for Astronomy and Astrophysics, MC 249-17 California Institute of Technology, Pasadena CA 91125, USA}
\email{kaitshin@caltech.edu}

\author[0000-0002-2088-3125]{Kendrick Smith}
\affiliation{Perimeter Institute for Theoretical Physics, 31 Carline Street N, Waterloo, ON N25 2YL, Canada}
\email{kmsmith@perimeterinstitute.ca}


\begin{abstract}
Massive galaxy clusters can introduce gravitational lensing and populations of suppressed star formation member galaxies into the line of sight, potentially changing the distribution of observable sources toward them from that of the average sky. As a result, Fast Radio Bursts (FRBs) aligned with galaxy clusters can provide a unique window into parameter spaces of the FRB population that other lines of sight do not afford. In this study, we demonstrate that the second CHIME/FRB baseband catalog contains FRBs emitted from within or behind clusters identified from the latest DECaLS data release; we isolate a sample of 26 FRBs where this is likely, including one repeating FRB and two FRBs that intersect the Coma cluster. Comparing our results against a range of simulated FRB populations, we conclude that the number of these associations represents a $3\sigma$ rate enhancement toward galaxy clusters, constituting $1.4\pm0.4\%$ of the FRBs detected in the second CHIME/FRB baseband catalog. We suggest that this enhancement is caused by an equal proportion of member galaxies hosting additional FRBs, and gravitational lensing magnifying background sources. We demonstrate that such contributions are sensitive to alternative progenitor channels and high redshift evolution in the FRB population, providing a future avenue for constraining these features. Considering the redshifts and masses of the associated clusters, we identify FRB\,20211113A, which is aligned with the strong gravitational lens Abell 2218 as a potential lensed candidate for further consideration.
\end{abstract}

\section{Introduction}\label{sec:intro}
Population models describing an ensemble of Fast Radio Burst (FRB) detections have, to date, relied on fairly simple treatments of the FRB population. Specifically, state-of-the-art population synthesis models such as those implemented in \texttt{z-DM} \citep{jamesFastRadioBurst2021} and other works \citep{shinInferringEnergyDistance2023, Jain2026}, often describe FRB energies following a single power law and assume that the FRB population traces cosmic star formation. These simple models successfully describe the first generation of FRB observations, with independent agreement between starkly different surveys obviating the need for further complexity \citep{jamesFastRadioBurst2021, shinInferringEnergyDistance2023, jamesModellingRepetitionZDM2023}. As more data on the FRB population are gathered, however, two important features should be noted. Firstly, in well-sampled parts of the observational parameter space, small departures from these simple models are beginning to emerge. Secondly, the dominance of a single FRB survey, CHIME/FRB (Canadian Hydrogen Intensity Mapping Experiment), in the detection of unique FRB sources leaves large areas of the observational parameter space undersampled, allowing for large departures from these models to go unnoticed. In both cases, any such departures could hold key information about the nature of FRB progenitors. 

An example of an emerging discrepancy can be found in observations of FRB host galaxies. Under the assumption that FRBs dominantly trace star formation, which has typically been consistent with observations \citep{gordonDemographicsStellarPopulations2023, loudasUnveiling2025}, the population of FRB hosts should comprise primarily star-forming galaxies. Recently, however, a demographic study of these hosts, by \citet{horowiczHostGalaxiesFast2026}, has revealed a population of low-star-formation host galaxies that cannot be reconciled with FRBs tracing only star formation, requiring up to $\sim$20$\%$ of FRBs to originate from a different progenitor that instead traces a longer time delay channel (e.g. one which is proportional to stellar mass). 

Where a high density of observed FRBs in the nearby universe reveals small discrepancies with population models, the insensitivity of the largest FRB survey to high-redshift FRB populations allows potentially large model departures to go unnoticed. To date, more than 95$\%$ of FRB sources have been discovered by CHIME/FRB, with only $\sim 15\%$ of those with energies $E\geq10^{39}$ ergs, expected to originate from redshifts $z\geq1$ \citep{shinInferringEnergyDistance2023}. As a result, any evolution in the population or rate turnover at higher redshifts has only a marginal impact on the detected population.

In either case, the observed impact of any departure from the simple models will be small and therefore washed out when averaged over the entire FRB population, restricting us from detecting these more complex features which could inform our understanding of FRB progenitors. To improve our sensitivity to these features, we must identify a subset of the FRB population for which these differences are exaggerated, such as lines of sight containing massive galaxy clusters. 

Galaxy clusters represent a relatively different galactic environment than the field, with star formation expected to be suppressed due to the high cluster temperatures \citep{laganaQuenching, giodiniStellarTotalBaryon2009, leauthaudIntegratedStellarContent2012} and ram pressure stripping \citep{ramPressureHester, ramPressureBoselli} of the member galaxies. The number of FRBs originating from within galaxy clusters, therefore, provides important information on the fraction of FRBs born from non-star-forming evolutionary channels. Galaxy clusters can also effect substantial magnification of background sources through gravitational lensing \citep{kneib_cluster_2011}. For observations along cluster sightlines, this increases sensitivity to high redshift sources \citep{bradac_focusing_2009}. As a result, any sight lines containing a massive cluster can also be more sensitive to high-redshift evolution in the energy function and source density of FRBs. 

Understanding the rate of FRBs from galaxy-cluster lines of sight can therefore serve as an insightful case study for understanding the progenitors of the FRB population. In this work, we use the second CHIME/FRB baseband catalog \citep{Shin2026} to prepare a sample of FRBs viable for population analysis (selected prescriptively from a uniform survey sample) that we expect to be emitted from within or behind galaxy clusters. We also demonstrate the response of this dataset to emerging discrepancies of interest, highlighting the general utility of FRB samples from within or behind galaxy clusters.

In \S 2, we outline our method for isolating a population of cluster FRBs within a catalog of unlocalized bursts. In \S 3, we apply this method to identify which of the CHIME FRBs are likely associated with known clusters and compare their properties with those of the broader population. In \S 4, we discuss the implications of these over-densities in the context of simulated expectations and draw conclusions about the contributions to FRB rates from cluster members vs cluster lensing.  In \S 5, we conclude, summarizing the key takeaways from the paper and suggesting future directions.

\section{Method}\label{sec:method}

Galaxy clusters are expected to affect the apparent detection rate of FRBs in three cardinal ways. Firstly, the member galaxies of a cluster are potentially sources of additional FRBs, and thus the introduction of a cluster into the line of sight causes an increase in the expected local FRB detection rate \citep{connorDeepSynopticArray2023a}. Secondly, a galaxy cluster can magnify faint background sources via gravitational lensing, causing otherwise sub-threshold events to become detectable, again causing an increase in the expected local detection rate \citep{sammonsForecastingFastRadio2025}. Finally, any multipath scattering or dispersion introduced by the intra-cluster medium (ICM) temporally broadens FRB signals \citep{macquartKoay, sammonsForecastingFastRadio2025}, making them more difficult to detect and decreasing the expected local detection rate. 

These effects are expected to be negligible when averaged over an entire FRB sample, and therefore, to quantify them, we must first isolate the subsample of FRBs where the relative impact of these effects is exaggerated, e.g. FRBs emitted from within or background to a galaxy cluster. Identifying this sample requires a metric indicating the likelihood of cluster association for a given FRB, and a catalog of FRBs large enough to reasonably contain cluster associations. For simplicity, we use a uniform sample of FRBs detected by only a single instrument, thereby avoiding consideration of the differential selection effects between FRB surveys. Specifically, we use the CHIME/FRB survey, as it contains the greatest number of unique FRB sources. 

CHIME is a wide-field-of-view radio interferometer located at the Dominion Radio Astrophysical Observatory in Penticton, British Columbia \citep{chime2018}. Its unique design comprises four 20$\times$100-m half cylinder reflectors, oriented along the N-S direction. This yields a primary beam with an approximately $\sim230$ square degree field of view in a narrow ellipse spanning $-11^\circ\leq \text{Dec}\leq110^\circ$ (reaching 20 degrees past the north celestial pole). Each cylinder is instrumented with 256 dual polarisation feeds operating in the $400-800\,$MHz band. From these inputs, the CHIME/FRB backend uses FFT (fast fourier transform) beamforming \citep{ngCHIMEFRBApplication2017} to synthesize 1024 tied-array beams with average FWHM (full width half maximum) of $\sim\lambda/D=21.5$', which are searched continuously for FRB signals. Using injections of simulated FRB signals, the CHIME/FRB detection pipeline is empirically estimated to be 90\% complete to FRBs above a 5\,Jy\,ms fluence threshold \citep{merryfieldInjectionSystemCHIME2023, McGregor2026}.

The recently released second CHIME/FRB catalog contains 4539 bursts, originating from 3641 unique sources distributed across the northern sky \citep{collaborationSecondCHIMEFRB2026a}. For all observed bursts, intensity dynamic spectra are saved, allowing for signal analysis at the native, $\Delta t=0.98304\,$ms \& $\Delta \nu=24.4\,$kHz, resolution of the FRB search and sky localization to $\sim$ degree precision. For a sub-sample of FRBs with S/N$\geq12$, raw voltage (baseband) data are saved for each antenna, allowing for high resolution analysis and localization to $\sim 1$ arcminute \citep{michilliAnalysisPipelineCHIME2021, CHIMEBasecat1}. For intensity-only bursts, the corresponding header localization areas \citep[see][]{CHIMERN1} are typically much larger than the characteristic angular size of massive galaxy clusters ($\sim$ arcminutes) and therefore we use only the subsample of the second CHIME/FRB catalog with baseband data, constituting the $\sim 1600$ FRB sources of the second CHIME/FRB baseband catalog \citep{Shin2026}. This sample represents the largest-ever number of FRB sources observed by a single instrument that are localized to the 1'$-$10' angular scale characteristic of massive galaxy clusters. 

To determine the probability of a cluster association for each FRB, we construct a heuristic "score" metric that is intended to maximize for cluster-associated FRBs. We then determine the probability of each FRB's score under the null hypothesis that any spatial coincidences between galaxy clusters and FRBs are chance alignments of background clusters with unassociated foreground FRB sources. To do so we evaluate the null hypothesis distribution using Monte Carlo simulations. 

To construct this score metric we begin by considering the angular offset of a FRB from known galaxy clusters. Depending on the source of cluster associations the expected distribution of their offsets will change. Lensed FRBs are likely to trace the regions of high magnification that lie at low offsets ($r\lesssim r_{500}$) \citep{johnsonLENSMODELSMAGNIFICATION2014}. Conversely, FRBs from member galaxies should trace the projected number density of galaxies within the cluster, which is described well by a Navarro, Frenk and White (NFW) profile with a concentration of $c=2.6$ \citep{navarroStructureColdDark1996, budzynskiRadialDist2012}. To balance these potential contributions we characterize the offset score $\Sigma_s$ using a projected NFW profile with a higher concentration of $c=6$, truncated to a maximum value of $\Sigma_s(r=0.1\,r_{500}/c)$ \citep{budzynskiRadialDist2012}.


To evaluate $\Sigma_s$ for each FRB we synthesize the all-sky map of $\Sigma_s$ as the sum of individual cluster profiles, normalising each to the $M_{500}$ mass of its halo, where $M_{500}$ is the mass enclosed within the $r_{500}$ radius within which the mean matter density is 500 times the Universe's critical density. For simplicity, we populate our map with a homogeneous sample of clusters from only one survey. Many widefield galaxy cluster surveys exist, constructed from a variety of cluster tracers such as X-ray luminosity, optical galaxy clustering, or amplitude fluctuations in the thermal Sunyaev-Zel'dovich effect. The largest of these samples is the recent galaxy cluster catalog released by \cite{wenCatalog158Million2024}, who derive the positions, masses, and redshifts for nearly 1.6 million clusters of galaxies identified in DECaLS observations (Dark Energy Camera Legacy Survey). This sample shares substantial overlap with the CHIME field of view, making it the natural choice for this work. This cluster sample is, however, incomplete, containing no galaxy clusters near the Galactic plane, and therefore to similarly avoid low Galactic latitudes we consider only the subset of the second CHIME/FRB baseband catalog survey footprint that lies within $20\times r_{500}$ of a detected cluster, yielding a masked sample of 892 FRBs valid for our analysis.

By considering the null distribution of $\Sigma_s$, it would be reasonable to isolate a subsample of FRBs likely to be associated with clusters as those least likely to be consistent with the null distribution. Foreground FRBs, however, would represent a substantial false positive signal within this sample, which we can ameliorate by excising from the considered sample, all bursts foreground to clusters. If our sample contained FRBs with high precision localizations ($\lesssim$ arcsecond) such a cut would be trivial as it would be routine to determine the host galaxy and infer a redshift. A drawback of using the CHIME/FRB sample, however, is that without the more recently implemented Outrigger array \citep{outriggersDesign, RBFLOAT2025}, the detections in the second baseband catalog have insufficient localization precision to allow robust host galaxy associations $\approx95\%$ of the time \citep{Andersen2026}\footnote{
We further caution that the PATH algorithm \citep[Probabilistic Association of Transients to their Hosts][]{aggarwalPATH} typically applied for host galaxy associations is not designed to account for the increased number density of potential host galaxies within galaxy clusters. In general, this effect will cause the confidence of PATH associations toward galaxy clusters to be biased high and so should be interpreted with care, even for high precision localizations. To correct for this, the luminosity function of cluster member galaxies should be used to estimate their probability of chance alignment compared to background galaxies.}. As a result, the vast majority of our FRBs lack a corresponding redshift. However, analysis of the ensemble population \citep{jamesZDMDistributionFast2022, shinInferringEnergyDistance2023, Jain2026} constrains the underlying population of FRBs, from which we can estimate the distribution of possible redshifts for an FRB detected by CHIME with a given extragalactic DM along a typical line of sight, $P(z|\text{DM})$. To derive a burst's extragalactic DM we subtract the expected Milky Way contribution from its observed DM, using the latest Galactic electron distribution model by \cite{ockerNE2025UpdatedElectron2026}. From this we can estimate the probability that an observed FRB was emitted foreground to the cluster, $P(z<z_{\text{clust}}|\text{DM})$, by integrating $P(z|\text{DM})$ from $z=0$ to the redshift of the cluster ($z_{\text{clust}}$). Specifically, we use the \texttt{z-DM} implementation of this calculation \citep{jamesZDMDistributionFast2022}, assuming a population set by the parameters in Table \ref{tab:inputs} from \cite{jamesMeasurementHubblesConstant2022}, with the exception of $\gamma$ and $n_{\text{SFR}}$, for which we test a range, as they are both uncertain and highly influential on the observed population. Given that clusters are expected to contribute a large DM to FRBs within or background to them, we expect $P(z<z_{\text{clust}}|\text{DM})$ to be low in the event of a true cluster association. 

Incorporating both $\Sigma_s$ and $P(z<z_{\text{clust}}|\text{DM})$ into our score metric, we evaluate the total score ($\mathcal{L}$) of a given FRB as the weighted average over the FRBs localization uncertainty region of $\Sigma_s$ divided by the probability that a given FRB is foreground to the most distant\footnote{To be conservative, we sum the densities of overlapping clusters and use the redshift of the most distant one.} cluster in the line of sight, i.e.,  $\mathcal{L}=\Sigma_s(\text{RA},\text{Dec})\,/\,P(z<z_{\text{clust}}|\text{DM})$. The resulting score is largest for high-DM FRBs aligned closely with nearby clusters, where we have the highest confidence that those FRBs are either hosted within or background to the cluster. Conversely, the score will be lowest for FRBs that are far from all clusters, where the cluster's matter density approaches that of the field. We also highlight that for an equivalent FRB impact parameter, the cluster association score will in general be lower for more distant clusters ($z\gtrsim1$) as we approach the telescope's detection horizon. 

To transform the score into a probability of cluster association, we characterize the distribution of $\mathcal{L}$ using Monte Carlo simulations of our observed FRB sample, under the null hypothesis that any spatially coincident galaxy clusters are chance alignments with unassociated foreground FRB sources (see Appendix \ref{app:montecarlo} for more details on this process). We then calculate the probability of chance coincidence ($P_{cc}$) for each FRB as the fraction of all Monte Carlo realisations with $\mathcal{L}\geq\mathcal{L_{\text{FRB}}}$. Significant cluster associations in our FRB sample will then manifest with $P_{cc}$ values that are inconsistent with draws from a uniform distribution between zero and one, given the sample size.

\begin{table*}
    \setlength{\arrayrulewidth}{0.5mm}
    \centering
    \begin{tabular}{|c|c|c|}
        \hline
        \textbf{Parameter} & \textbf{Input Value} & \textbf{Description}\\
        \hline
        $E_{\text{max}}$  & $10^{41.7}$ erg & Maximum FRB energy\\
        $E_{\text{min}}$ & $10^{30}$ erg & Minimum FRB energy\\
        $n_\text{sfr}$  & 1 & Cosmic star formation rate scaling index \\
        $\alpha$ & -1.03 & Spectral Rate index \\
        $\gamma$  & -1 & Energy function index \\
        $\mu_\text{Host}$ & 2.23 & Mean of the normal $\log_{10}$DM$_{\text{Host}}$ distribution in pc cm$^{-3}$\\
        $\sigma_\text{Host}$ & 0.57 & Standard deviation of the normal $\log_{10}$ DM$_{\text{Host}}$ distribution in pc cm$^{-3}$\\
        $\Phi_0$ & $2.43\times10^4$ Gpc$^{-3}$yr$^{-1}$& Volumetric FRB rate at $z=0$\\
        \hline
    \end{tabular}
    \caption{Input parameters for baseline \texttt{z-DM} population forecast. All parameters are identical to those used by \citet{sammonsForecastingFastRadio2025}, inherited from the best fits of \citet{jamesMeasurementHubblesConstant2022} using flat priors on all parameters, with the exception of $\Phi_0$, which we set to approximately yield the observed CHIME/FRB rate as described by \citet{sammonsForecastingFastRadio2025}, $E_{\text{max}}$ as determined by \citet{ryderLuminousFastRadio2023}, and both $\gamma$ and $n_{\text{SFR}}$ for which a range of values are tested. Other parameters not listed here retain their default values that can be found in the public \texttt{z-DM} implementation and in the analysis of \citet{jamesMeasurementHubblesConstant2022}.}
    \label{tab:inputs}
\end{table*}

\section{Results \& Discussion}\label{sec:results}
Using the above method, we can calculate the probability that each of the 892 FRBs in our sample is foreground to the most distant cluster with a projected offset $\leq3\times r_{500}$, which is expected to contain all lensing action and 99\% of member galaxies \citep{budzynskiRadialDist2012, johnsonLENSMODELSMAGNIFICATION2014}. As we are using $P(z|\text{DM})$, the resulting $P_{cc}$ value depends on the population model assumed for the FRB sample, with the most impactful parameters being the cumulative slope of the energy function $\gamma$ and the shape of the redshift distribution, $\rho(z)$, which is often parametrized as the cosmic star formation rate to the $n$th power $\rho(z)=\psi(z)^{n_{\text{SFR}}}$. Best estimates of these parameters to date still have wide error bars, $\gamma=-0.3^{+0.8}_{-0.5}$, $n_{\text{SFR}}=1.72^{+1.48}_{-1.10}$ \citep[see Appendix B]{shinInferringEnergyDistance2023}, $\gamma=-0.95^{+0.18}_{-0.15}$, $n_{\text{SFR}}=1.13^{+0.49}_{-0.41}$ \citep{jamesMeasurementHubblesConstant2022}, $\gamma=-1.16^{+0.57}_{-0.68}$, $n_{\text{SFR}}=0.91^{+0.61}_{-0.55}$ \citep{hoffmannModellingDSAFAST2024a} allowing for a range of potential $P(z|\text{DM})$ distributions. To provide a reasonable $P(z|\text{DM})$, we assume the best-fit population of \citet{jamesMeasurementHubblesConstant2022}, but calculate results for $\gamma=\{0.0,-0.5,-1.0\}$ and $n_{\text{SFR}}=\{0.5,1.0,1.5\}$. 

\begin{figure*}[t]
    \centering
    \begin{minipage}{0.49\textwidth}
        \centering
        \includegraphics[width=\linewidth]{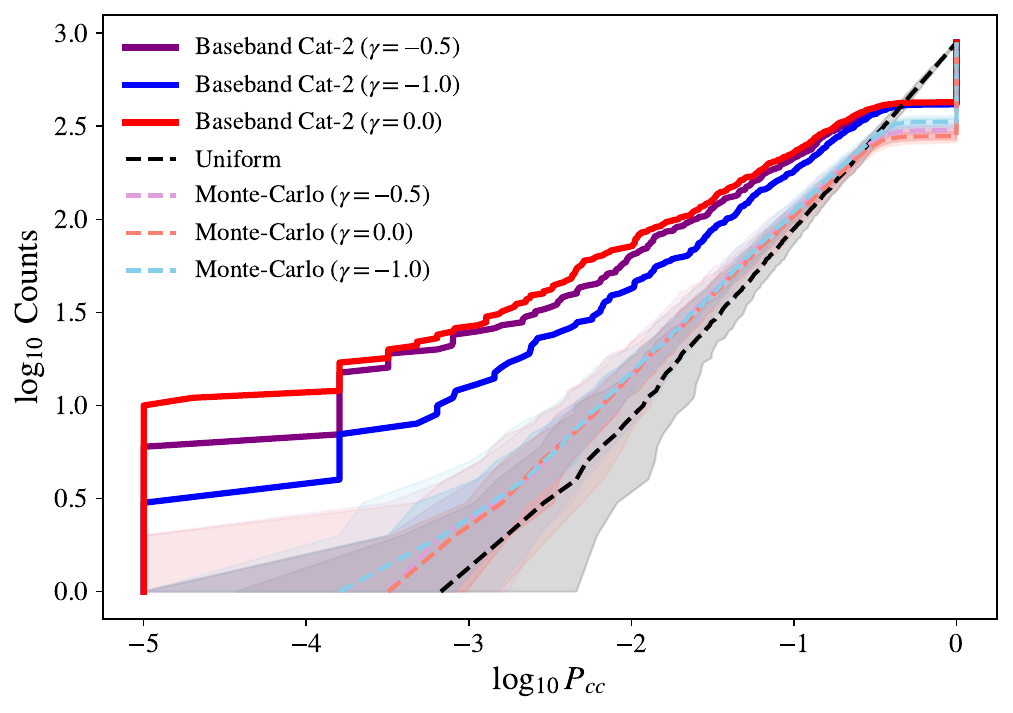}
    \end{minipage}
    \hfill
    \begin{minipage}{0.49\textwidth}
        \centering
        \includegraphics[width=\linewidth]{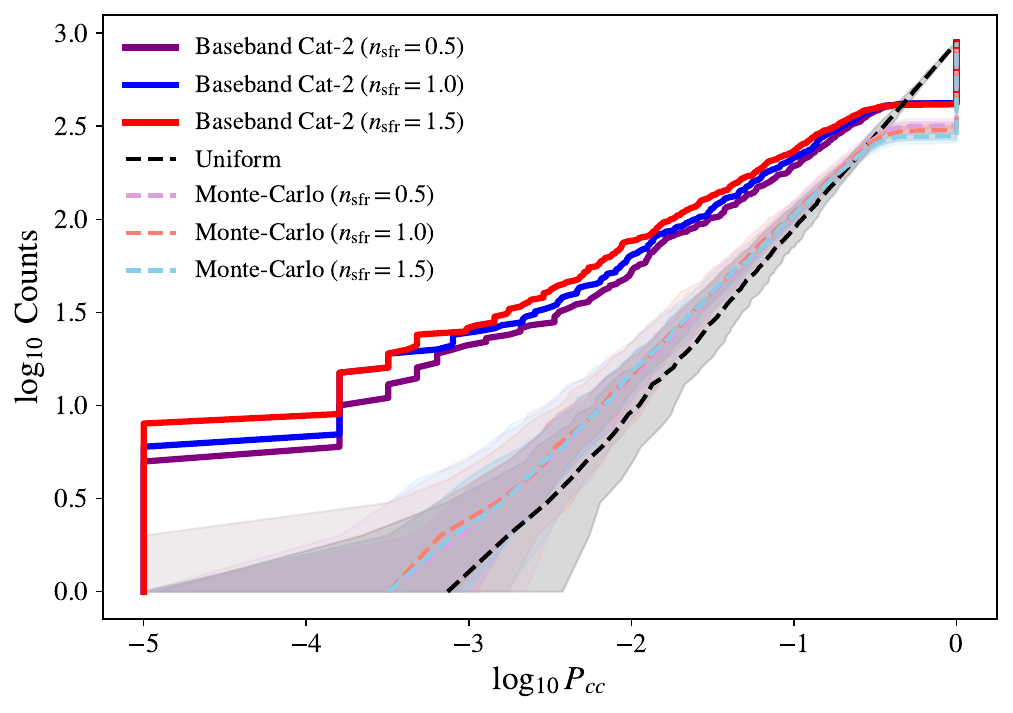}
    \end{minipage}
    \caption{Cumulative distributions of the probability of chance coincidence ($P_{cc}$) of FRB samples under the null hypothesis that all FRBs are emitted foreground to the most distantly aligned cluster. Plotted distributions correspond to the FRBs from the second CHIME/FRB baseband catalog that lie within the DECaLS footprint and to associated Monte Carlo realisations under the null hypothesis, in comparison to realisations of a true uniform distribution. \textit{Left:} These distributions under variation in the slope of the power-law FRB energy function  $\gamma=0$, $\gamma=-0.5$, $\gamma=-1$. \textit{Right:} These distributions under variation in the FRB redshift distribution to $\rho(z)\propto\psi(z)^n$, where $\psi(z)$ is the cosmic star formation rate \citep{madauCosmicStarFormation2014}. In all cases, the data show a clear excess in low $P_{cc}$ events compared with Monte Carlo simulations of the null hypothesis, with Anderson-Darling tests confirming that each baseband catalog distribution has a probability of less than $p\ll0.001$ of being drawn from the corresponding Monte Carlo distribution. From this we conclude that the CHIME/FRB sample must contain FRBs background to DECaLS clusters.}
    \label{fig:baseCat2Probs}
\end{figure*}

These resulting $P_{cc}$ CDFs are provided in Fig. \ref{fig:baseCat2Probs}, with the different values of $\gamma$ and $n_{\text{SFR}}$ in the left and right panels, respectively. An Anderson-Darling test of the second CHIME/FRB baseband catalog sample against the Monte Carlo distributions yields $p\ll 0.001$ for all combinations plotted in Fig. \ref{fig:baseCat2Probs}. Therefore, regardless of the specific population values, we can reject the null hypothesis that all FRBs are foreground to the most distant cluster within $3\times r_{500}$ at high significance. Instead, we conclude that some FRBs within the second CHIME/FRB baseband catalog must pass through or be hosted within the clusters in the \cite{wenCatalog158Million2024} sample. Additionally, Fig. \ref{fig:baseCat2Probs} shows that the $P_{cc}$ for a given FRB is dependent on $\gamma$. This dependence is expected, as shallower indices make higher energy events more likely, skewing $P(z|\text{DM})$ to higher redshifts and decreasing $P_{cc}$ for any given FRB. Conversely, the dependence of $P_{cc}$ on $n_{\text{SFR}}$ is weaker, with high $n_{\text{SFR}}$ values similarly skewing $P(z|\text{DM})$ and marginally decreasing $P_{cc}$. As a compromise between the best-fit CHIME/FRB and CRAFT $\gamma$ values \citep[][respectively]{shinInferringEnergyDistance2023, jamesMeasurementHubblesConstant2022}, we will proceed using a fiducial $\gamma=-0.5$ and explicitly consider the uncertainty in $\gamma$ where relevant. As the specific value of $n_{\text{SFR}}$ is less impactful, we will proceed under the assumption that $n_{\text{SFR}}=1$, as we find it to be the most physically meaningful choice. 

\begin{figure}
    \centering
    \includegraphics[width=0.6\linewidth]{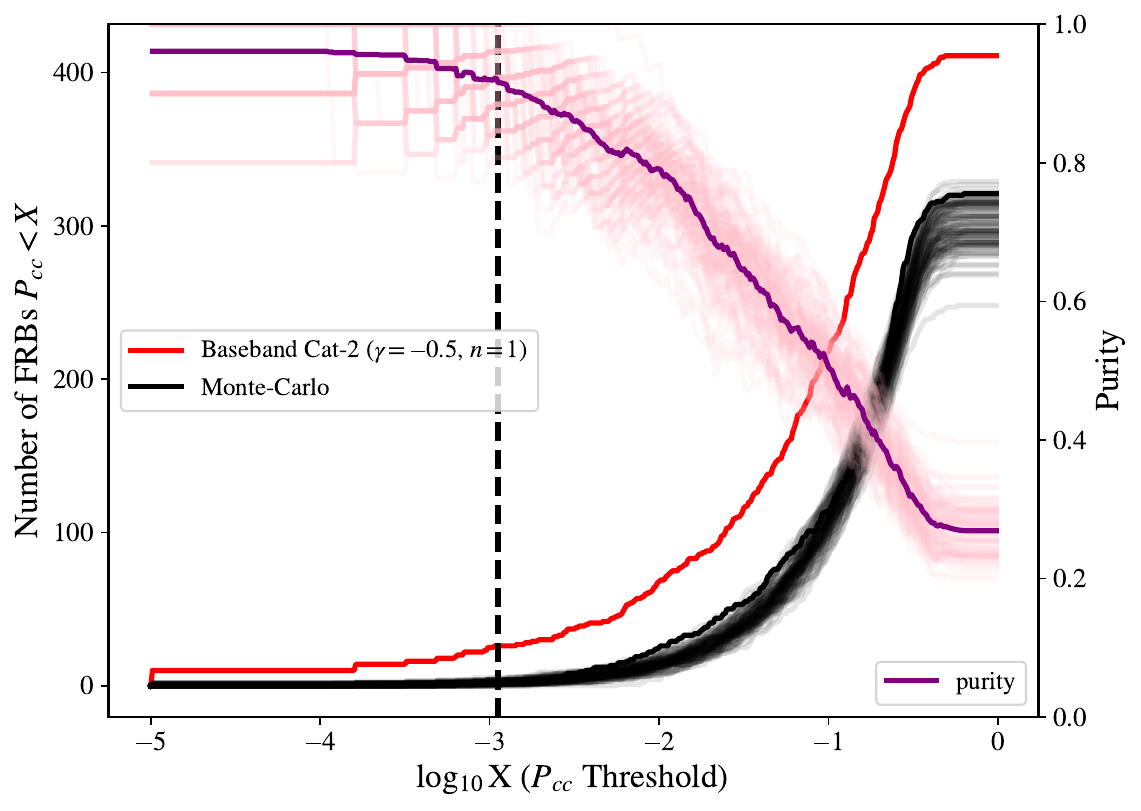}
    \caption{Size of the FRB subsample satisfying $P_{cc}<X$ as a function of $X$ for both the second baseband catalog (red), and 100 independent Monte Carlo realisations (black). Also shown is the purity of the baseband catalog 2 sample when compared against the false positive rate in each Monte Carlo realisation (pink) and averaged over all realisations (purple). The dotted line marks where $X=1/N$ where $N$ is the size of the unmasked FRB sample ($N=892$).} 
    \label{fig:samplePurity}
\end{figure}

To isolate a subsample of FRBs for which cluster associations are the most likely, we perform a cut on $P_{cc}$, including all those FRBs satisfying $P_{cc}\leq X$, where $X$ is an arbitrary threshold in the range $[0,1]$. Figure \ref{fig:samplePurity} shows the purity of this subsample (true positive candidates / total candidates) with respect to the Monte Carlo realisations of the null hypothesis as a function of $X$. Figure \ref{fig:samplePurity} shows that the purity and size of the subsample vary monotonically with $X$, and therefore $X$ may be chosen arbitrarily to optimize the dataset's utility in a specific context. In our context we are interested in balancing contamination with completeness, and therefore we set $X=1/N$ where $N$ is the size of the unmasked FRB sample ($N=892$). This yields a subsample of 26 FRBs with $P_{cc}<1/N$ detailed in Table \ref{tab:FRBs} with a purity of $92\pm5\%$ (mean and standard deviation of the pink curves seen in Fig. \ref{fig:samplePurity}), i.e., we expect $2.2\pm1.2$ false positive bursts that satisfy our criteria but are actually foreground to their aligned galaxy clusters. This false positive rate is slightly higher than the mean expected false positive rate of 1 for uniform random variables ($P_{cc}$). We expect that this is caused by the finite Monte Carlo sampling and sharp truncation of the halo profiles at $r\geq 3r_{500}$. Despite its mild departure from expectation, the false positive rate remains low and as such does not prohibit the interpretation of this dataset. 

\subsection{Cluster Sample Properties}
We now compare the properties of our isolated sample of cluster FRBs to those of the broader FRB sample used in the analysis to identify any morphological differences. We highlight here that while cluster FRBs are expected to have a large extragalactic DM, far in excess\footnote{We highlight one counterpoint from \cite{lanmanConstrainingBaryonFractions2025}, where FRB 20230703A, intersecting three lower mass groups, has a lower than expected extragalactic DM for its redshift} of predictions for bursts originating at the same redshift \citep{connorDeepSynopticArray2023a, rafiei-ravandiStatisticalAssociationCandidate2024, lanmanConstrainingBaryonFractions2025}, we may not compare DM distributions between the cluster and non-cluster FRB samples derived here. This is because the DM was used to calculate the cluster association score $\mathcal{L}$, which was used to select the cluster sample, and is therefore biased in that sample. We investigate the dependence of DM on cluster offset in an unbiased way in \S \ref{subsec:DMprofile} below.

Apart from an excess extragalactic DM, FRBs intercepting galaxy clusters and groups are also expected to have comparable scattering times to the average FRB population but may present with broader intrinsic pulse widths, resulting from scattered rise times, imparted by multiple environments of comparable scattering, being interpreted as intrinsic pulse structure \citep{sammonsForecastingFastRadio2025}. As a subsample with a higher chance of intercepting massive galaxy clusters, these 26 FRBs represent a chance to test these hypotheses. To do so, we use the burst morphology measurements derived from fitting on the intensity data provided with the second CHIME/FRB catalog \citep{collaborationSecondCHIMEFRB2026a}. In part this choice is due to the availability of morphological measurements from the second CHIME/FRB baseband catalog. It is, however, also motivated by potential biases in fitting results from baseband data of cluster populations introduced by the loss of high frequency ring buffer data at the high DMs more prevalent in cluster populations \citep{chime2018}. Conversely, the intensity data retain all bandwidth, but cannot be coherently dedispersed and so suffer instead from uncompensated intra-channel DM smearing. This smearing is accounted for in the applied fitburst method \citep{fonsecaFitburst}, and therefore we find this to be the appropriate dataset for our purposes. 

\begin{figure}
    \centering
    \includegraphics[width=\linewidth]{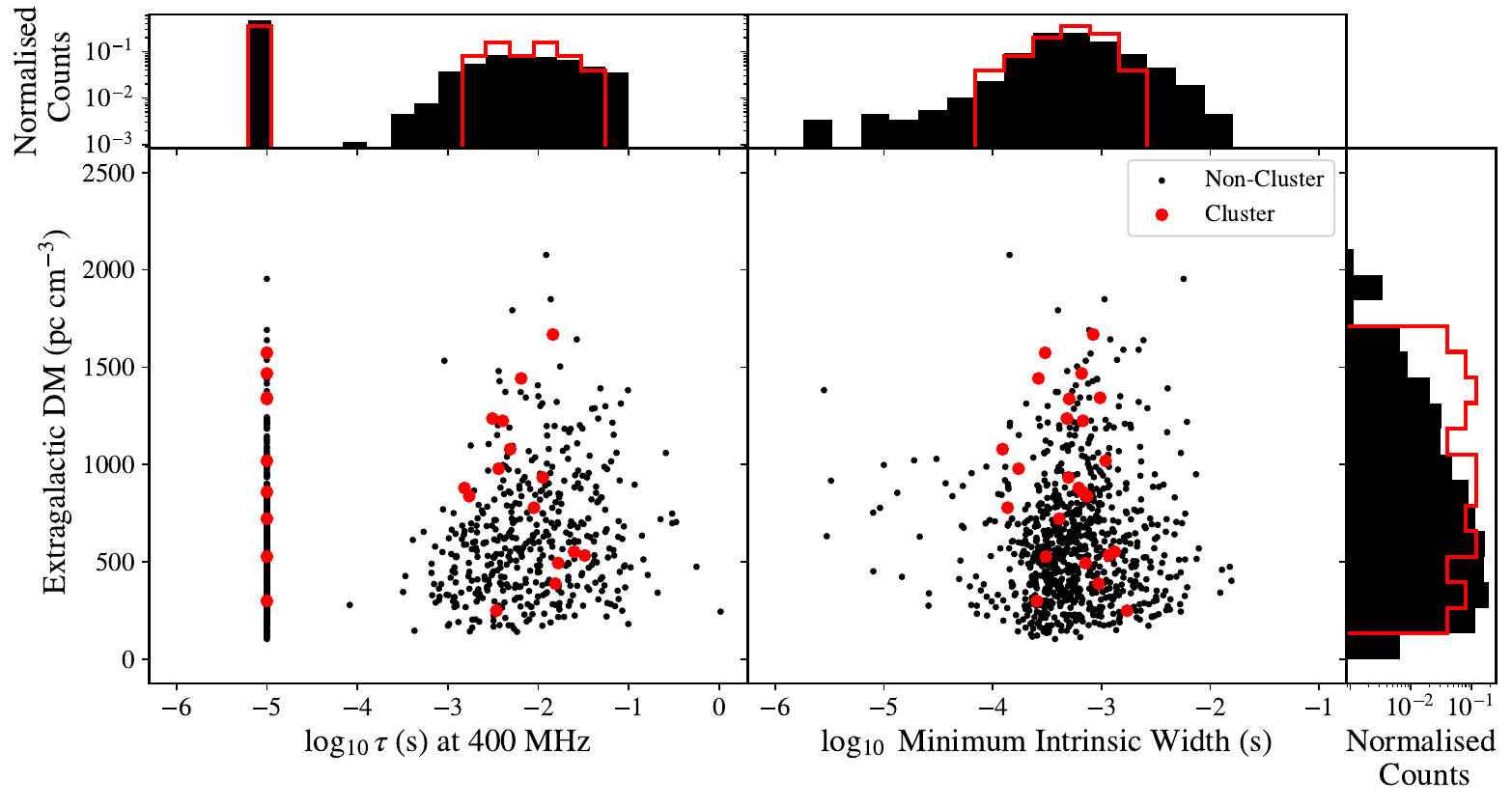}
    \caption{Comparison of FRB morphology amongst cluster and non-cluster FRB samples. Specifically, we display extragalactic DM as a function of both scattering time (\textit{left}) and intrinsic width ($\textit{right}$) estimated from the intensity fits published in CHIME/FRB catalog 2 \citep{collaborationSecondCHIMEFRB2026a}. FRBs for which a scattered profile was disfavoured are binned at $\tau=10^{-5}\,$s for visual distinction. Marginalized distributions show that all quantities except DM are consistent between populations.}
    \label{fig:DMvsMorphology}
\end{figure}

Fig. \ref{fig:DMvsMorphology} shows the DM, scattering time ,$\tau$, referenced to 400\,MHz, and the minimum intrinsic width of all burst components\footnote{A burst component here refers to a distinct peak or structure in the FRBs dynamic spectra} in the second CHIME/FRB catalog, as fit from the intensity data using fitburst \citep{collaborationSecondCHIMEFRB2026a}. The Figure depicts these measures for both the cluster and non-cluster subsamples, marginalizing each into normalized histograms on the periphery. In the case of scattering times, where the model fitting preferred an unscattered profile over a scattered one, we set $\tau$ to 0.01\,ms\footnote{This is intended to represent that any scattering was too small to be detectable. An important caveat, however, is that the minimum detectable scattering time is a function of intrinsic width and S/N, and therefore may be significantly different from 0.01\,ms}. 

Following \cite{sammonsForecastingFastRadio2025}, our expectation is that FRBs traversing galaxy clusters pick up a secondary scattering contribution from a log-normal distribution $\log_{10}\tau \sim \mathcal{N}(-2.2,0.7)$\,s at 400\,MHz, which is coincidentally similar to the observed scattering distribution of non-cluster FRBs. The convolution of two exponential scattering kernels causes the smaller of the two scattering times to present instead as a rise time, which may lead to cluster FRBs with a wider apparent intrinsic-width distribution. As a result, we expected our cluster and non-cluster samples to display consistent scattering times, and inconsistent minimum intrinsic widths, with the cluster sample having larger intrinsic widths approximately described by $\log w_{\text{int}}\sim\mathcal{N}(-2.2,0.7)$\,s. For each marginalized distribution, we test the hypothesis that cluster and non-cluster samples are drawn from the same underlying population using both Anderson-Darling and Kolmogorov-Smirnov tests. We find that both the minimum width and scattering times of the cluster and non-cluster distributions are consistent with being drawn from the same population ($p\geq0.25$). This conclusion lies counter to our expectations, and suggests that the scattering times contributed by clusters are less than $\mathcal{N}(-2.2,0.7)$ distribution we expected, potentially smaller than $\log_{10}\tau \sim \mathcal{N}(-3.2,0.3)$\,s, which approximately describes the distribution of intrinsic widths for the cluster sample.

Suppressed Galactic scintillation due to angular broadening is another characteristic behaviour expected to manifest in FRBs traversing clusters \citep{connorDeepSynopticArray2023a, sammonsForecastingFastRadio2025}. Scintillation has been positively identified within a number of CHIME/FRB baseband detections \citep{nimmoScint, curtinScint, RBFLOAT2025, Fine2026}, however to confidently evaluate a lack of scintillation in observed FRBs requires a precise investigation of CHIME/FRB's ability to measure narrow frequency modulation, and distinguish scintillation from other sources of spectral variation such as self-noise. Progress towards this goal is underway within the CHIME/FRB collaboration, and thus we leave remarks on these properties to a future work.



\subsection{Cluster DM Contribution}\label{subsec:DMprofile}
As discussed above, the DMs in our cluster sample are biased high\footnote{If we isolate a sample using the method in \S \ref{sec:method}, FRBs at larger cluster separations will have DMs that are biased high to meet an equivalent likelihood threshold as lower impact parameter FRBs. As such, the above sample of cluster associations should not be used to infer the mean cluster DM profile.} by the search method and should not be used to infer DM contributions from the clusters themselves. To investigate the cluster DM response in a less biased way, we instead isolate a sample of likely cluster interceptions by evaluating only the probability a FRB is foreground to its nearest cluster given its DM and the redshift of the nearest DECaLS cluster, $P(z<z_{\text{clust}}|\text{DM})$, regardless of its separation. FRBs with a low probability of being foreground to their nearest cluster can then be used to construct a subsample that is a tracer of any universal cluster DM profile. 

\begin{figure}
    \centering
    \includegraphics[width=\linewidth]{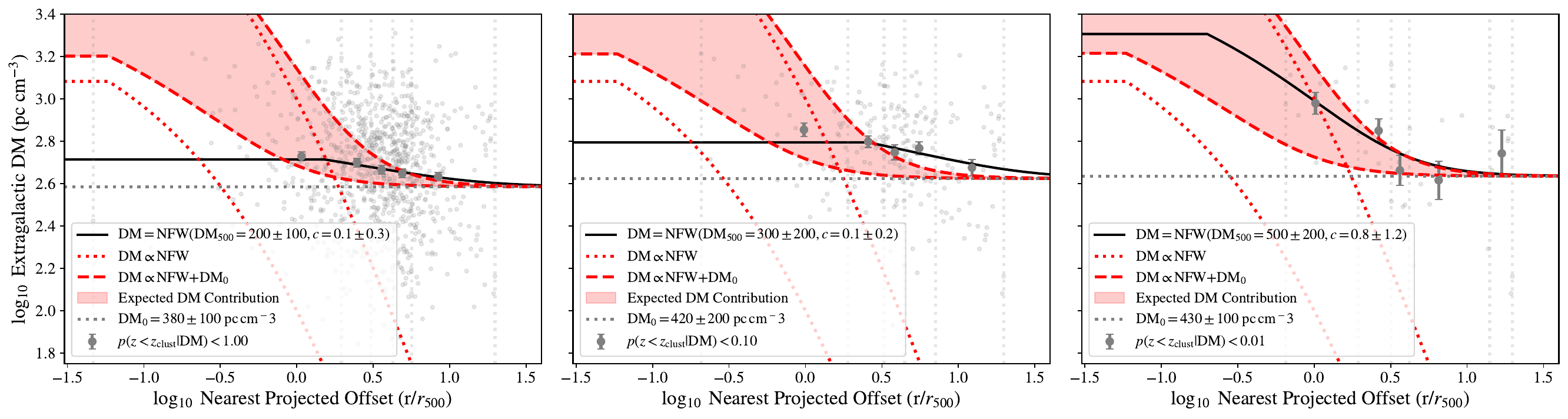}
    \caption{Extragalactic DM as a function of physical impact parameter normalized by the $r_{500}$ scale radius of nearby galaxy clusters measured by \citep{wenCatalog158Million2024}. DM mean and variance as a function of impact parameter, shown by the dark grey points, are calculated from five equally populated bins demarcated by the dashed lines, these points reflect the mean of the binned impact parameter values as well as DM, and therefore do not lie in the centre of each bin. Each panel shows the results for a subsample of the FRBs satisfying $P(z<z_{\text{clust}}|\text{DM})<P_x$, for $P_x$=\{1.0,0.1,0.01\}, with the purity of cluster associations in the subsample increasing from left to right respectively. In each case the black lines represent the best-fit projected NFW profile truncated to a maximum at 0.1 scale radii, with DM at $r_{500}$ (DM$_{500}$), concentration (c) and DM offset (DM$_0$) as free parameters. The red regions highlight the range of DM contributions typically expected for galaxy clusters, based on the range of electron density profiles derived from X-ray observations of known clusters \citep{cavagnoloIntraclusterMediumEntropy2009}. Specifically, the dashed lines bound NFW profiles with DM$_{500}$=100-1000 pc\,cm$^{-3}$, $c=2.6$ and DM$_0$ set to best fit value for that panel. For higher purity cluster samples the data show a greater mean DM fluctuation and a response curve in closer agreement with the range of expected NFW profiles suggesting that these DM contributions originate from galaxy clusters.}
    \label{fig:DMProfile}
\end{figure}

Fig. \ref{fig:DMProfile} visualizes this for several subsamples satisfying $P(z<z_{\text{clust}}|\text{DM})<P_{\text{th}}$, for decreasing foreground probability thresholds $P_{\text{th}}=\{1.0,0.1,0.01\}$. In each panel we show the extragalactic DM as a function of impact parameter normalized by the $r_{500}$ radius of the nearest cluster. For comparison, Fig. \ref{fig:DMProfile} also highlights characteristic DM contributions given by projected NFW profiles with $c=2.6$, truncated to a maximum density at 0.1 scale radii (qualitatively similar to typical cored $\beta$-profiles with a steeper drop off at high radii \citep{betaModels}). Informed by X-ray observations \citep{cavagnoloIntraclusterMediumEntropy2009}, we normalize these characteristic profiles to contribute a DM = 100-1000 pc\,cm$^{-3}$ at $r=r_{500}$ over the blank field mean. For the complete sample ($P(z<z_{\text{clust}}|\text{DM})<1.0$), the data show a significant correlation between impact parameter and DM, indicating that clusters are contributing a detectable DM to the FRB population. The weak dependence of mean DM on impact parameter, however, suggests that the DM contribution from clusters represents only a small fraction of the total DM averaged over the detected FRB population. Fitting a similar truncated NFW profile to this data yields a best fit concentration parameter of $c=0.1\pm0.3$, well below the characteristic expectation. As $P_{\text{th}}$ decreases, corresponding to an increase in the purity of cluster FRBs, the average DM contribution from clusters is enhanced and a more obvious DM profile emerges in agreement with the characteristic expectation for $P(z<z_{\text{clust}}|\text{DM})<0.01$. This suggests that DM contributions from clusters are indeed present in our sample and that the mean DM cluster profile may be approximately represented by a truncated NFW form with a concentration of $c=0.8\pm1.2$ and a DM at $r=r_{500}$ of DM$_{500}=500\pm200\,$pc\,cm$^{-3}$.

Another common observable for galaxy clusters is the amplitude of the thermal Sunyaev-Zel'dovich effect \citep{zeldovichThermal}, which is proportional to the product of electron density ($n_e$) and temperature ($T$) integrated over the line of sight to the CMB ($y_{\text{sz}}\propto \int_0^{d_{\text{CMB}}}  n_eT\,d\ell$). As discussed by \cite{connorDeepSynopticArray2023a} and performed by \cite{takahashiMeasurementAngularCrosscorrelation2025}, the tandem use of DM and $y_{\text{sz}}$ for a given line of sight can be used to estimate the gas temperature for intervening clusters. To do so robustly requires localization of the FRBs to host galaxies with known redshifts, such that the DM contributions from the intergalactic medium may be subtracted. As we lack the localization precision to accurately identify host galaxies, we cannot confidently measure cluster gas temperatures with individual bursts. At the population level, however, a universal cluster profile will introduce a relation between $y_{\text{sz}}$ and DM for cluster FRBs, which we may interpret as mean ICM temperature under the following assumptions: the observed $y_{\text{sz}}$ is dominated by the path length that is foreground to the FRB source, and the line-of-sight temperature profile is uniform. 

\begin{figure}
    \centering
    \includegraphics[width=\linewidth]{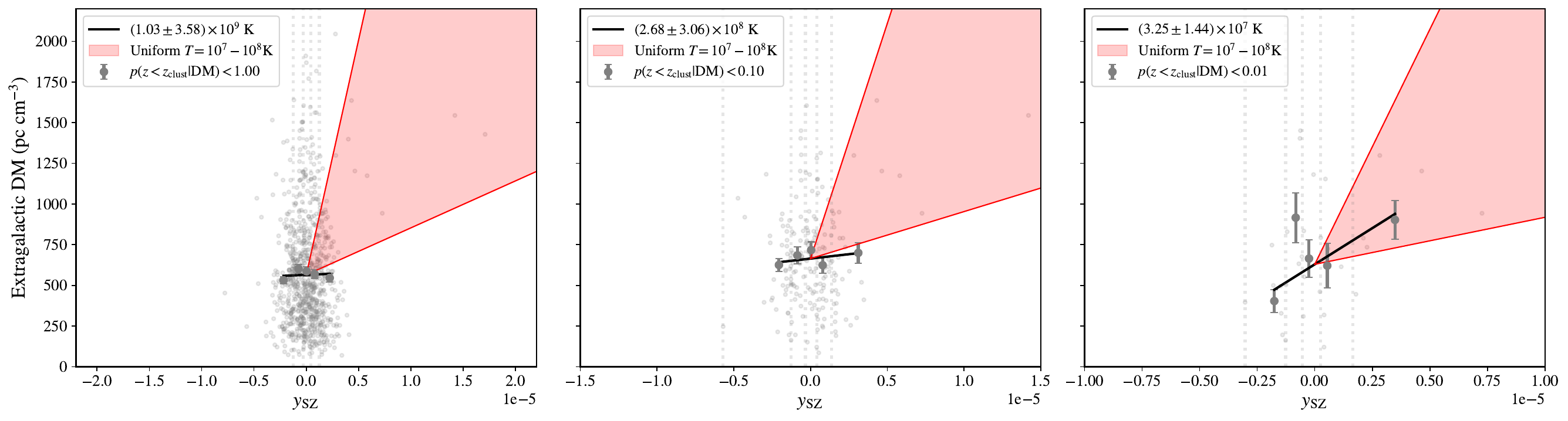}
    \caption{Extragalactic DM as a function of $y_{\text{sz}}$ calculated from the Planck map of the thermal Sunyaev-Zel'dovich effect as the weighted mean over the localization region \citep{adePlanck2015Results2016}. DM mean and variance as a function of $y_{\text{sz}}$, shown as dark grey points, are calculated from five equally populated bins demarcated by the dashed lines, these points reflect the mean of the binned $y_{\text{sz}}$ values as well as DM, and therefore do not lie in the centre of each bin. Each panel shows the results for a subsample of the FRBs satisfying $P(z<z_{\text{clust}}|\text{DM})<P_x$, for $P_x$=[1.0,0.1,0.01], with the purity of cluster associations in the subsample increasing from left to right respectively. In each case, the black lines are the best-fit linear relations between $y_{\text{sz}}$ and DM. In the regime where changes in DM and $y_{\text{sz}}$ are dominated by a common path length along the line of sight, such as the path through the ICM, the gradient of the relation will represent the average temperature of electrons along that path, under the assumption of a uniform temperature profile. The red regions highlight the expected range of electron temperatures for the ICM of massive clusters, 10$^{7}$--10$^8\,$K. The alignment of higher purity cluster samples with the typical temperatures expected for the ICM reinforces that the excess DM seen in Fig. \ref{fig:DMProfile} is contributed by the ICM of galaxy clusters}
    \label{fig:tempProfile}
\end{figure}

Using the same unbiased sample as above, Fig. \ref{fig:tempProfile} shows FRB extragalactic DM as a function of $y_{\text{sz}}$ for an increasingly pure sample of cluster FRBs. For the entire sample ($P(z<z_{\text{clust}}|\text{DM})<1.0$), any relation between DM and $y_{\text{sz}}$ becomes washed out by sources foreground to known clusters, violating the underlying assumptions and causing a spuriously high temperature to be inferred. As the purity increases a majority of FRB sources become background to known clusters (provided that emission from member galaxies is rare, see \S \ref{subsubsec:members}) and the temperature becomes more meaningful. In this regime the inferred temperature of $T_{\text{ICM}}=3.25\pm{1.44}\times10^7\,$K falls into the characteristic range $10^7-10^8\,$K expected for galaxy clusters \citep{cavagnoloIntraclusterMediumEntropy2009}.

In combination with the DM profiles suggested by Fig. \ref{fig:DMProfile}, these results are consistent with an average contribution to FRB DMs from a hot intracluster medium following a universal NFW profile. While not highly constraining, these results demonstrate that the nature of clusters may be measured with FRBs even without robust host localizations. Furthermore, they highlight that while the DM contribution from clusters is only large in a handful of FRBs, it is detectable at the population level and therefore may introduce a significant source of DM error for precision cosmological studies. We leave a more in-depth treatment of cluster DM contributions to a future endeavour.

\subsection{Specific Case Studies}
Amongst the identified sample of cluster associated FRBs, there are several examples of specific interest; one especially high mass cluster candidate, two FRBs piercing the well-known Coma cluster at different radii and one established repeating FRB \citep{pleunisRN3,cookRN4}. We visualize these cases in figures \ref{fig:out1}--\ref{fig:out3} below, with each figure displaying the FRB dynamic spectra synthesized from the baseband data and the burst localization with respect to the cluster. In the dynamic spectra, we choose an arbitrary resolution to optimize visual distinction. We highlight that most signals are missing high-frequency portions of their bandwidth as a result of high-DM FRBs exceeding the length of the CHIME/FRB voltage ring buffer. In the sky localization panels, we visualize the corresponding galaxy clusters in a number of ways, contouring the $y_{\text{sz}}$ parameter in the region around the cluster from the Planck map, outlining the $r_{500}$ radius of each cluster derived by \cite{wenCatalog158Million2024} in white, and highlighting the galaxy cluster members within $r_{500}$, as derived by \citet{wenCatalog158Million2024}, as colored circles.

None of the FRBs we consider have associated host galaxies. We deliberately make no attempt to apply the PATH algorithm and identify a host for these FRBs due to the low localization precision of the data. As discussed in \S \ref{sec:method}, despite the improvement afforded to localizations by using the baseband data, there is a only a  $\sim$5\% chance that a secure PATH association can be made for any FRBs in the second CHIME/FRB baseband catalog \citep{Andersen2026}. Moreover, the PATH algorithm is not designed to account for the increased number density of candidates within galaxy clusters that make chance interceptions of host candidates more likely at the cluster redshift. A full accounting of these effects is beyond our scope, and therefore we leave an evaluation of host associations in galaxy clusters to a future work. In the following subsections we briefly discuss each of the interesting cases.

\subsubsection{FRB\,20211113A}\label{subsec:out1}
\begin{figure}
    \centering
    \includegraphics[width=0.7\linewidth]{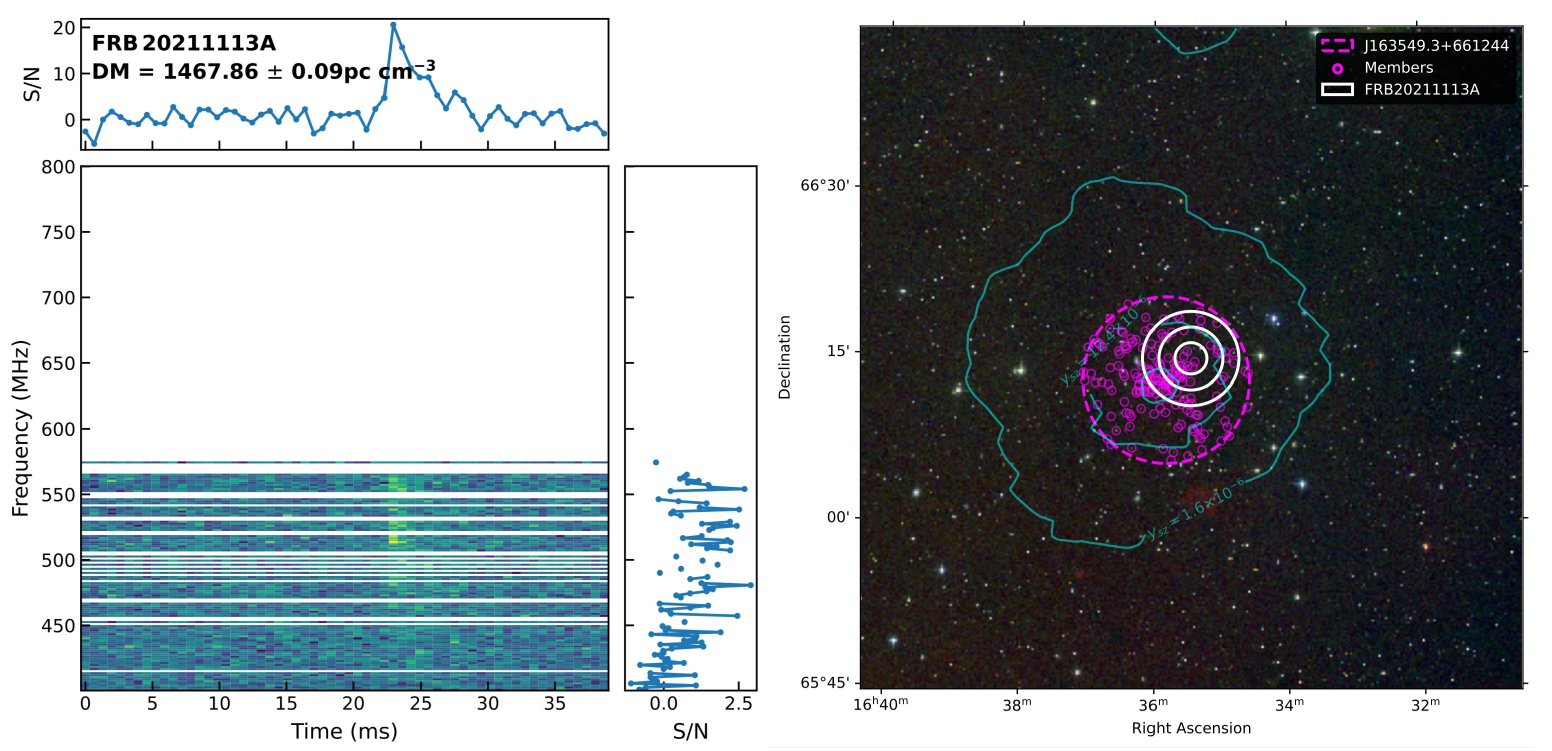}
    \caption{CHIME/FRB observation of FRB\,20211113A. \textit{Left}: Dynamic spectrum of the FRB dedispersed to the DM listed in the top left. As noted in the text, much of the high frequency bandwidth is missing, due to baseband ring buffer losses of early times at high dispersive delays. \textit{Right}: Burst line of sight from DSS2 imaging \citep{galDigitizedSecondPalomar2004} with the corresponding $y_{\text{sz}}$ values from \cite{adePlanck2015Results2016} contoured in cyan. The $r_{500}$ scale radii of aligned galaxy clusters from \cite{wenCatalog158Million2024} demarcated via a dotted circle and associated member galaxies from \cite{zouPhotometricRedshiftsGalaxy2022, liPhotometricRedshiftCatalogue2024} highlighted as small circles in the same color. Finally the 1,2, and 3$\sigma$ localization uncertainty regions for the FRB are shown in full white lines.}
    \label{fig:out1}
\end{figure}

FRB\,20211113A intersects the highest mass cluster amongst our cluster associated FRBs, with an $M_{500}$ from DECaLS of $\sim9.2\times10^{14}\,M_\odot$. As seen in Fig. \ref{fig:out1}, several member galaxies fall within the localization ellipse for this burst, allowing for the possibility that it originates from within the cluster itself. Despite being a more distant cluster amongst our sample, J163549.3+661244 (more commonly known as Abell 2218) is still relatively nearby, with $z_{\text{cluster}}\approx0.18$, being far below the notional maximum of $z_{\text{max}}=2.0$, corresponding to $P(z<z_{\text{max}}|\text{DM})=0.999$. As a result, if the FRB host is a member of the galaxy cluster, the host and cluster must contribute a joint excess exceeding $1000$\,pc cm$^{-3}$, which is approximately consistent with the large $y_{\text{sz}}=1.7\times10^{-5}$ value at the burst location for an ICM with $T\approx 5\times10^7\,$K. 

Conversely, Abell 2218 is also a known strong gravitational lens, capable of large magnifications \citep[$\mu\approx26$; ][]{kneibMagni}. The burst localization overlaps the cluster's strong lensing region within $2\sigma$, opening the door for FRB\,20211113A to be strongly lensed. To delineate between these possibilities, we consider the relative rates between member galaxy emission and cluster lensing in \S \ref{subsec:forecasting}.

\subsubsection{FRB\,20210205B \& FRB\,20210411D}\label{subsec:out2}
\begin{figure}
    \centering
    \includegraphics[width=\linewidth]{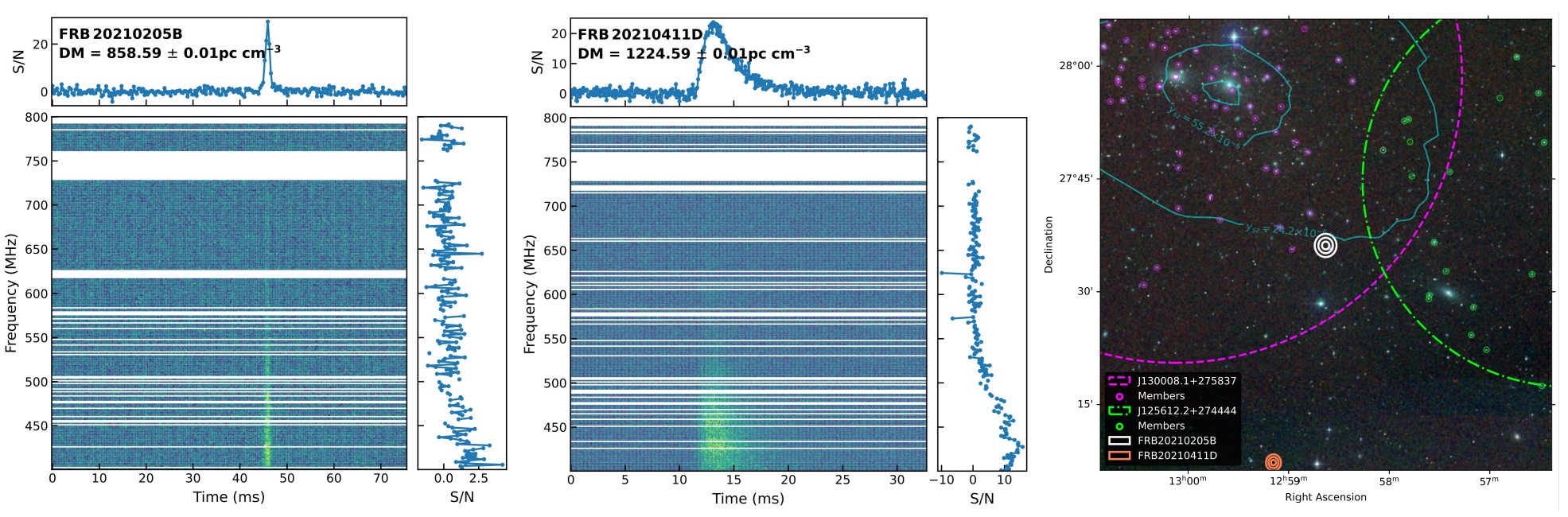}
    \caption{As in Fig. \ref{fig:out1} but for FRB\,20210205B $\&$ FRB\,20210411D.}
    \label{fig:out2}
\end{figure}

FRBs\,20210205B \& 20210411D are both aligned with the most well studied cluster amongst our sample, offset $\approx0.8\,$Mpc and $\approx1.4\,$Mpc respectively from the centre J163549.3+661244, known more commonly as the Coma Cluster. Fig. \ref{fig:out2} visualizes these alignments, highlighting both the Coma and Leo galaxy clusters, which are themselves members of the Coma supercluster \citep{gregory1978}. The proximity of the Coma cluster at a comoving distance of $\approx120\,$Mpc \citep[$z\approx0.027$, assuming a Planck cosmology][]{planck18} necessitates that if either FRB is embedded within the cluster, $\gtrsim90\%$ of its total DM (858.59\,pc\,cm$^{-3}$ and 1224.64\,pc\,cm$^{-3}$ respectively) must be contributed by the host galaxy and cluster contributions. Conversely, if the cluster and host contribute negligibly, the hosts could be as distant as $z\approx1.3$ and $z\approx1.8$. Despite being more closely aligned with the cluster, FRB\,20210205B shows a markedly lower DM than FRB\,20210411D, indicating that the cluster may not be dominating the DM of both bursts. FRB\,20210205B shows negligible scattering and a relatively low fluence, whereas FRB\,20210411D has a measured scattering time of $4.0\pm0.1\,$ms referenced to 400\,MHz and is limited to higher fluences of $F_\nu\gtrsim11\pm3$ Jy\,ms. These properties provide little constraint on the distance to the source, however, we note that for a shallow energy function (cumulative $\gamma\leq-1.5$) such as that expected for FRBs \citep{jamesMeasurementHubblesConstant2022, shinInferringEnergyDistance2023}, lower fluence bursts are expected to more commonly originate at greater distances, suggesting that the host galaxy DM contribution could be dominating the DM of FRB\,20210411D. Finally, a Coma member galaxy lies within 2$\sigma$ of FRB\,20210205B. While we cannot definitively associate the FRB to the member galaxy, emission from within the cluster could lead to a smaller cluster DM contribution, potentially explaining the smaller DM of this burst compared to FRB\,20210411D. 

\subsubsection{FRB\,20201125B}\label{subsec:out3}
\begin{figure}
    \centering
    \includegraphics[width=0.8\linewidth]{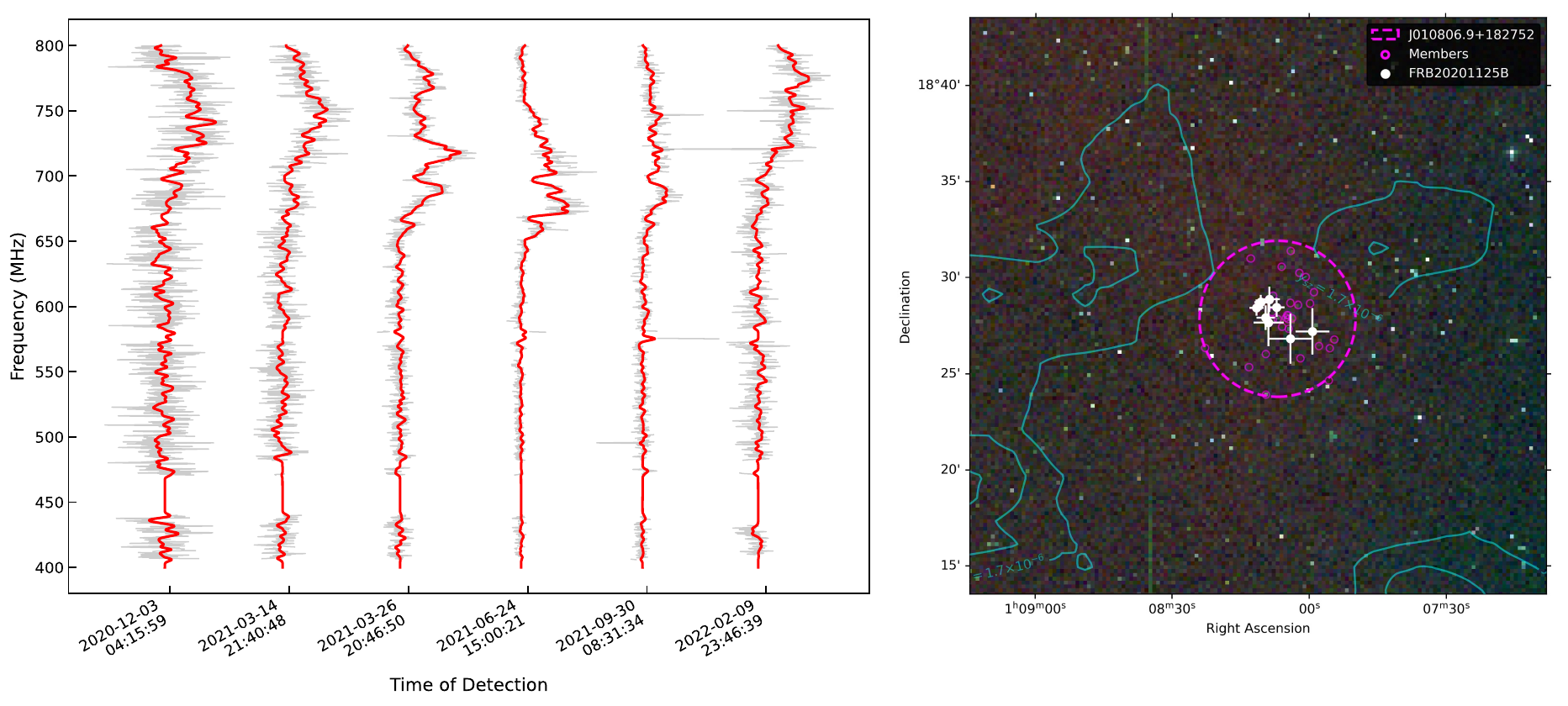}
    \caption{Arrival times and baseband localizations for all repetitions of FRB\,20201125B. \textit{Left}: The S/N of the time integrated burst spectra as a function of arrival time. \textit{Right}: Burst localizations with respect to aligned clusters, as in above figures.}
    \label{fig:out3}
\end{figure}

FRB\,20201125B is the only FRB in our sample belonging to a repeating source, namely FRB\,20200929C, with eight known repetitions in the second CHIME/FRB baseband catalog\footnote{FRB\,20201125B is the first burst from this source for which baseband data was collected}. Fig. \ref{fig:out3} shows the spectra of these repetitions over time and the spatial localization of each burst with respect to the cluster J010806.9+182752. This alignment was previously noted by \cite{ibikRepeater}, who also proposed that the cluster could be contributing the DM of the source. As a repeating FRB behind a cluster, FRB\,20200929C provides a unique opportunity to constrain any evolution or fluctuation in small scale properties of the ICM over long timescales in the future. Furthermore, it presents the best opportunity to date to find a strongly gravitationally lensed repeating FRB from which precision measurements on $H_0$ can be made \citep{ZhengStrongly2018}. The time delays associated with multiple imaging by a galaxy cluster lens can range from months to years with angular separations on the order of 1--10'' \citep{kelly2015Sci, Rodney2021NatAs,Pierel2024ApJ,Pascale2025ApJ}. Given the common bandwidths between some of the bursts, it is possible that the observed activity already contains gravitationally lensed copies of individual repetitions. The mass of the cluster is however a factor of a few smaller than other observed galaxy cluster lenses $\gtrsim 5\times10^{14}M_\odot$, making strong lensing by the cluster less likely. While unlikely, the potential utility of such a source mandates that any potential candidates be considered seriously, particularly because the echoed spectra, from which a lensed FRB can be identified, can easily be confused for common repeating FRB morphologies \citep{curtinMorph2025}. Moreover, the variable magnifications between images will correspond to variable energy detection thresholds, causing each image to appear as a distinct source with only some bright bursts in common due to the power-law distribution of repeating FRB bursts rates with energy \citep{ouldBoukattineRepeater2026}. This will further complicate the identification of lensed repeating FRBs from burst arrival times \citep{ZhengStrongly2018}, particularly for unknown lenses. Independently of these effects, specialized very long baseline interferometers (VLBI), such as the CHIME/FRB Outriggers and future CHORD/FRB Outriggers, provide a way to verify lensing by localising each observed burst to sub-arcsecond precision. This allows the spatial separation of images to be evaluated, and lensing confirmed. We leave a complete description of the phenomenology of strongly lensed repeating FRBs, and the VLBI localization of this repeater to a future work, but encourage follow-up observations of this source during active periods \citep{cookRN4}.

\subsection{Rate Expectations}\label{subsec:forecasting}
The Monte Carlo method described in \S \ref{sec:method} allows us to identify which FRBs are likely to originate from within or behind a galaxy cluster, but it does not provide any information regarding how many FRBs we should expect to identify this way. Broadly, these FRBs will fall into one of two categories: either they will be unmagnified background FRBs that happen to be aligned with a foreground cluster, which we term interlopers, or they will be additional FRBs detected only due to the presence of the cluster, constituting a cluster rate excess. 

To evaluate the number of expected interlopers we simulate 10000 mock FRB samples using \text{z-DM} and distribute them on the sky, uninformed by the location of known clusters, consistently with the spatial distribution expected for CHIME/FRB, as detailed in Appendix \ref{app:montecarlo}. As a baseline we assume the population parameters given in Table \ref{tab:inputs}, and evaluate all rate estimates at the central frequency of $600\,$MHz, with a Gaussian synthesized beam profile \citep{michilliAnalysisPipelineCHIME2021} of FWHM$\sim\lambda/d=$21.5' and detection threshold of $5\,$Jy ms \citep{merryfieldInjectionSystemCHIME2023}. To evaluate the impact of our largest uncertainty, corresponding to $\gamma$, we evaluate all forecasts for each $\gamma=\{0,-0.5,-1.0\}$. We then repeat our search, synthesising entirely new Monte Carlo distributions for each mock catalog, as in \S \ref{sec:method}, deriving 10000 mock cluster associated samples for each case. All mock searches are performed assuming $\gamma=-0.5$, making the results directly comparable to those in \S \ref{sec:results}, but allowing us to constrain the uncertainty introduced by the wrong choice of $\gamma$.

The resulting number of interloper events are consistent with Poisson expectations, with $13.4\pm3.7$ for $\gamma=-0.5$, and $14.1\pm3.7$ and $11.6\pm3.4$ for $\gamma=0.0$ and $\gamma=-1.0$ respectively. The similarity of these results across $\gamma$ is expected due to the relatively small difference in predicted redshift distributions for CHIME/FRB detections between these $\gamma$ values, i.e., $P(z<z_\text{clust}|\text{DM})$ exhibits only small changes for $\gamma$ values in this range. Where the method in \S \ref{sec:method} strongly rejects the null hypothesis that all observed FRBs are foreground to the most distantly aligned cluster, these simulations characterize how many cluster associations should be expected when clusters do not impact FRB detection rates. Comparing with the 26 associations we detected in \S \ref{sec:results}, we find that the probability that all our associations are interlopers, is $p=0.0008$ for $\gamma=-0.5$, $p=0.0016$ for $\gamma=0$ and $p\leq 0.0001$ for $\gamma=-1.0$. We therefore conclude that regardless of the precise population modeling, our associations represent a $3\sigma$ detection of a rate enhancement toward galaxy clusters that is responsible for  $1.4\pm0.4\%$ of FRB detections in the second CHIME/FRB baseband catalog.

Given the complexity of this analysis, it is reasonable to question whether the above result is robust to other choices made during the analysis. The major choices are the concentration of the NFW profiles composing the $\Sigma_s$ map, the population model assumed when weighting the score by $1/P(z<z_{\text{clust}}|\text{DM})$, the $P_{cc}$ threshold chosen in \S \ref{sec:results} to identify cluster associated events and the assumed z-DM distribution of the simulated interloper events used to calculate the rate excess. Varying the $P_{cc}$ threshold we find that the significance of the excess stays approximately constant for thresholds $P_{cc}\leq1/N=1/892$, but decreases at higher thresholds as the number of interloper events rapidly increases and the search loses sensitivity. As shown above, the result is relatively insensitive to changes in the energy index, $\gamma$, of simulated interloper population. Extending this, we find that the result is similarly insensitive to large changes in other parameters: $n=n\pm0.5$, $\mu_{\text{host}}$=${0.3\mu_{\text{host}}-3\mu_{\text{host}}}$, $\sigma_{\text{host}}$=${0.3\sigma_{\text{host}}-3\sigma_{\text{host}}}$ and the completeness of CHIME/FRB to low DM bursts \citep{merryfieldInjectionSystemCHIME2023,McGregor2026}, with the probability of 26 interlopers being $p\leq$0.001 in each independently tested case.

Of the above choices, those related to the score metric can have a notable impact on the significance of any rate excess, with steeper assumed $\gamma$ values for $P(z<z_{\text{clust}}|\text{DM})$, yielding fewer events (as seen in Fig. \ref{fig:baseCat2Probs}) and lower profile concentrations for $\Sigma_s$ resulting in a greater percentage of interlopers. However, as the observed excess is extracted from the difference between results of the same search on both the simulated and observed populations, any change to significance that results from changing the search itself represents a change in the sensitivity of the search, rather than the validity of search results. Therefore, we find that the apparent excess is robust to the modeling choices made herein.

\subsubsection{Modeling the Rate Enhancement}
We expect that the enhancement in the rate of FRBs towards galaxy clusters may be explained by the emission of additional FRBs from member galaxies and the magnification of faint background sources. To evaluate that expectation, we calculate how each galaxy cluster identified by \cite{wenCatalog158Million2024} alters the DM and rate of FRB detections via both additional sources in member galaxies and through lensing contributions. Using these rates we recreate the mock samples above, deriving new mock cluster associations that include the predicted impact of galaxy clusters for a range of populations. Below we describe in detail how we model the rate contributions from member galaxies and lensed background sources, as well as how direct member contributions are expected to respond to alternative FRB formation channels, and lensing contributions, to redshift evolution.

To determine the rate contribution from galaxy cluster members, we first consider how the detected rate is related to the density of source objects in the Universe. Typically, the density of source objects is assumed to be separable within the rate equation as
\begin{align}
    R &= \int dz\, \Phi(z) \xi(z)\\
    \Phi(z) &= \Phi_0\,\phi(z)\, \Omega\,\frac{dV(z)}{dz},
\end{align}
where $\phi(z)$ is the normalized volumetric rate density of FRB sources corresponding to a rate of $\Phi(z)$ sources within the comoving volume element $dV(z)/dz$ subtending the survey solid angle $\Omega$. $\Phi_0$ is the volumetric rate density at $z=0$, and $\xi(z)$ includes all other parts of the rate integrand (survey sensitivity, cosmology, FRB energy function, etc) and may be interpreted as the mean fraction of the observable energy function at redshift $z$ given the survey sensitivity. The detected rate contribution from a galaxy cluster can then be calculated by setting $\Phi(z)=\delta(z-z_{\text{cluster}})\Phi_{\text{cluster}}$, where $\Phi_{\text{cluster}}$ is the number of FRB sources within a cluster at $z_{\text{cluster}}$.

Estimating the number of FRB sources within a cluster, however, is a more complex task, as galaxy cluster members are qualitatively distinct from field galaxies, generally exhibiting lower levels of star formation. A common hypothesis for FRB population synthesis is to assume that FRB source density traces star formation. While this hypothesis appears to accurately describe a majority of the FRB population \citep{jamesFastRadioBurst2021, gordonDemographicsStellarPopulations2023}, emerging evidence has highlighted that a mixed model, in which some fraction of the FRB population instead traces stellar mass, is required to explain the distribution of FRB host galaxies \citep{bhandariHostGalaxiesProgenitors2020, gordonMappingSpatialDistribution2025, horowiczHostGalaxiesFast2026}. The fractions of the FRB population tracing stellar mass ($f_\odot$) and the star forming efficiency of massive cluster members relative to their field counterparts ($\epsilon$) are both generally uncertain. We begin, therefore, by considering the number of FRB sources that these member galaxies would have in the unphysical limiting case where FRBs trace only star formation ($f_\odot=0$) and cluster galaxies are identical to field galaxies ($\epsilon=1$). 

From observed empirical relations \citep{giodiniStellarTotalBaryon2009, leauthaudIntegratedStellarContent2012, kravtsovStellarMassHalo2018}, we can estimate the total stellar mass of a given cluster from its halo mass ($M_{500}$) as $(M_\star/10^{14}\,M_\odot)=9300\,(M_{500}/10^{14}\,M_{\odot})^{0.6}$. Using the assumption of $\epsilon=1$, we can then estimate the total star formation rate of a cluster from its total stellar mass via a notional star formation efficiency of the average cosmic field ($\varepsilon$), modulated by a relative star formation efficiency in the cluster compared to the average $\epsilon$. To determine $\varepsilon$, we use the ratio of cosmic star formation rates and stellar masses from \cite{madauCosmicStarFormation2014}:
\begin{align}
    \psi(z) &= \psi_0\frac{(1+z)^{2.7}}{1+[(1+z)/2.9]^{5.6}}\,M_\odot\text{year}^{-1}\text{Mpc}^{-3}\\
    \rho_\star &= \rho_{\star,0}\int\limits_z^\infty \psi(z')\frac{z'}{H(z')(1+z')}\\
    \varepsilon(z) & = \frac{\psi(z)}{\rho_\star(z)},
\end{align}
where $\psi(z)$ is the cosmic star formation rate, $\rho_\star(z)$ is the cosmic stellar mass density, $\psi_0$ and $\rho_{\star,0}$ are the stellar mass and star formation rate densities at $z=0$ respectively, and $H(z)$ is the Hubble expansion rate. From the above, we can then calculate the star formation rate in the cluster as $\varepsilon(z) M_\star$, which under the assumption of $f_\odot=0$, yields the number of FRB sources within the galaxy cluster as 
\begin{equation}\label{phiclust}
    \Phi_{\text{cluster}}(z)=9300\,\frac{\Phi_0}{\psi_0}\,\epsilon\,\varepsilon(z)\,(M_{500}/10^{14}\,M_{\odot})^{0.6}\,,
\end{equation}
assuming that the cluster falls within $\Omega$ for a CHIME/FRB synthesized beam. 

\begin{figure}
    \centering
    \includegraphics[width=0.8\linewidth]{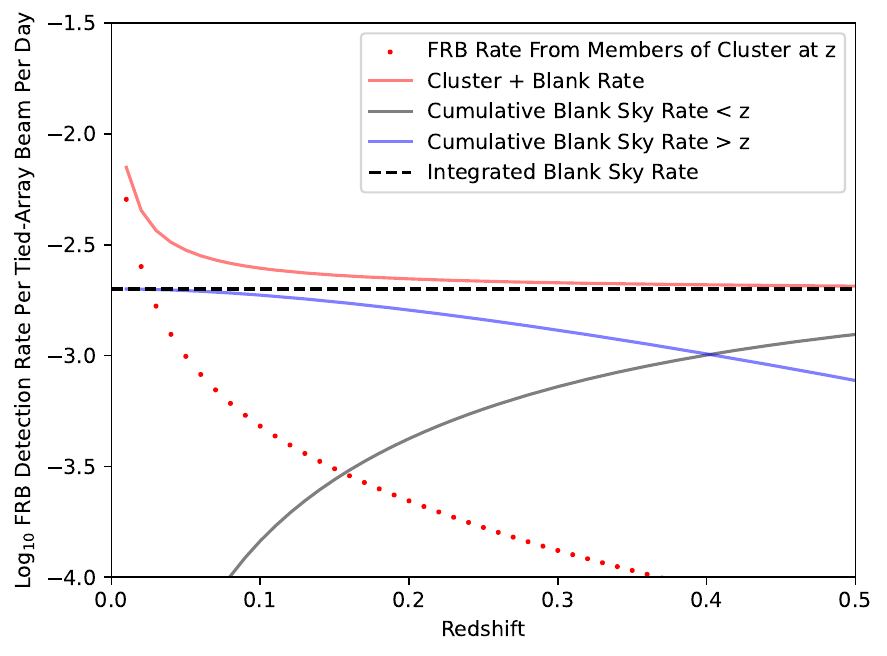}
    \caption{Expected rate contribution from the member galaxies of a $M_{500}=10^{15}\,M_\odot$ galaxy cluster compared to the blank field as a function of the clusters redshift. Rates are evaluated for the characteristic case where star formation efficiency in galaxy clusters is equivalent to that in the field, i.e., $f_\odot=0$, $\epsilon=1$, and $\gamma=-0.5$. The black dashed line shows assumed redshift-integrated CHIME detection rate per beam per day. The blue and grey lines decompose this rate into components above and below a given redshift. The red dotted line depicts the direct cluster contribution to the observed rate at a range of potential cluster redshifts and the full red line adds this cluster only contribution to the background rate.}
    \label{fig:memberContribution}
\end{figure}

Evaluating Eq. \ref{phiclust} for a massive galaxy cluster $M_{500}=10\times10^{14}\,M_\odot$, and our fiducial population ($\gamma=-0.5$), yields Fig. \ref{fig:memberContribution}. This figure shows that the direct contribution to the detected FRB rate from galaxy cluster members is strongly dependent on the cluster redshift, with only the most nearby ($z\lesssim0.025$) massive galaxy clusters expected to dominate over the contributions from a typical blank field in the unphysical $\epsilon=1$ regime. Given the linear dependence of the cluster contribution on $\epsilon$ weighed against the sharp rise in rate contribution, we also expect that massive galaxy clusters closer than $z<0.025$ will contribute comparably to the blank sky FRB detection rate for $\epsilon<0.1$. Finally, we caution that the cluster rates will also depend on the slope of the energy function, $f_\odot$ and $\epsilon$, e.g. if instead, $\gamma=-1.0$, the contribution from the galaxy cluster will dominate over the blank sky rate out to $z\lesssim0.1$.

To determine the impact of DM and lensing contributions on the background rate, we turn to the forecasting framework developed by \cite{sammonsForecastingFastRadio2025}. The framework is described more fully therein, but here we provide a brief primer. The method expands the rate calculation implemented in the \texttt{z-DM} package developed by \cite{jamesZDMDistributionFast2022}, to include the impact of gravitational lensing and DM contributions from massive foreground galaxy clusters by using their observed lens models \citep{johnsonLENSMODELSMAGNIFICATION2014, coeRELICSReionizationLensing2019} and electron density profiles \citep{cavagnoloIntraclusterMediumEntropy2009}. As a result, we can both calculate the distribution of FRBs expected to be detected by a given FRB survey (functionality native to the \texttt{z-DM} package), and predict how these detection rates will change as a function of DM and redshift for sources background to a variety of massive galaxy clusters.

Using the direct and lensed rate estimates in tandem, we can then calculate how each cluster in the \cite{wenCatalog158Million2024} sample contributes to the CHIME/FRB detection rate. To make this computationally feasible, we restrict ourselves to the most massive galaxy clusters as they are likely to have the highest impact. Specifically, we consider only direct member contributions from galaxy clusters, $M_{500}\geq 10^{14}\,M_{\odot}$, and for lensing $M_{500}\geq 5\times10^{14}\,M_\odot$, which is also restricted by the availability of lens models. As a result the ensemble impact of lensing on the observed FRB catalog will be underestimated, but we expect this effect to be small due to the reduced effectiveness of lower mass lenses. 

These rates can then be used to redraw mock catalogs of FRB redshifts and DMs consistent with the sample that we use from the second CHIME/FRB baseband catalog, including member galaxy and lensing contributions from the observed sample of galaxy clusters. For each mock catalog we can then re-perform the cluster association method from \S \ref{sec:method} to derive a sample of mock cluster associations as above. Finally, by trialling different population models, the number of cluster associations that we expect to detect can be evaluated over a range of scenarios and compared against our true population for physical insight. 

As discussed in \S \ref{sec:intro}, the scenarios that we are most interested in exploring are those where the FRB population comprises a mix of both stellar mass and star-formation-tracing progenitor evolution channels as suggested by \cite{horowiczHostGalaxiesFast2026}, and those where the FRB population evolves at high redshifts where our surveys are mostly insensitive. To model multiple progenitor channels in one population, we introduce a mixture model, where FRB source volume density tracks both stellar mass and star formation in proportions controlled by $f_\odot$ following
\begin{equation}
    \Phi(z) = (1-f_\odot)\Phi_{\psi,0}\frac{\psi(z)}{\psi_0}+\,f_\odot\Phi_{\rho,0}\frac{\rho_\star(z)}{\rho_{\star,0}},
\end{equation}
corresponding to a number of FRB sources within a cluster
\begin{equation}
    \Phi_{\text{cluster}}(z) = (1-f_\odot)\,\frac{\Phi_{\psi,0}}{\psi_0}\epsilon\,\varepsilon(z) M_\star+f_\odot \frac{\Phi_{\rho_{\star},0}}{\rho_0}M_\star\,,
\end{equation}
where $\Phi_{\psi,0}$ and $\Phi_{\rho,0}$ are number densities of FRB sources per star formation rate and stellar mass respectively. Rather than normalising the density of FRB sources at $z=0$, we choose to normalize these components to equalize $\int dz\, \Phi_\psi \xi(z) = \int dz\, \Phi_{\rho_{\star}} \xi(z)$ (i.e., to yield an equivalent number of FRBs detected FRBs in the field regardless of $f_\odot$). 

Conversely, to model evolution in the population towards high redshifts, we apply a rudimentary step change at $z=1$ in the baseline population of Table \ref{tab:inputs} to different values of the most impactful population parameters, $\gamma$ and $n_{\text{SFR}}$, normalized such that the distribution of detections in a blank field still changes smoothly as a function of redshift.

Further detail on the z-DM sampling can be found in Appendix \ref{app:rates}, however, as in the previous mock samples we assume a baseline population given by the parameters in table \ref{tab:inputs}. To explore the impact of mixed progenitor channels and redshift evolution, we trial the population ranges $f_\odot=\{0,0.2,0.4,0.6,0.8,1.0\}$, $\gamma'=\{-2.5,-2,-1.5,-1.0,-0.5\}$ and $n'_{\text{SFR}}=\{0,0.5,1.0,1.5,2.0\}$, where prime symbols denote the parameters of the evolved population (e.g. $\gamma$ at low redshift and $\gamma'$ at high redshift). To characterize the mean and variance in the expected number of cluster associations, we generate 100 mock samples of cluster FRBs for each trial population.

Within each mock sample, the dominant source of cluster associations comes from chance interceptions of unmagnified background FRBs with foreground clusters. On average the interlopers account for 13-14 cluster associations, consistent with the cluster independent simulations as expected. There is also little to no change in the number of interlopers over the trial populations in $f_\odot$, $\gamma'$ and $n'_{\text{SFR}}$, with the largest sensitivity being to $f_\odot$, which loses 1-2 interlopers on average as $f_\odot$ increases from 0 to 1, due to stellar mass tracers concentrating all FRB sources at lower redshifts. This relative independence of chance interceptions from evolution and mixed populations is consistent with our expectations, as these variations should only impact FRBs magnified by or emitted from within galaxy clusters. 

\subsubsection{Member Contributions}\label{subsubsec:members}
Like all galaxies, the members of a galaxy cluster are potential FRB hosts. Unlike their field counterparts however, the members of a galaxy cluster are often immersed in a hot ICM which may suppress the star formation required for the evolution of some FRB progenitors \citep{laganaQuenching, jamesFastRadioBurst2021}. In the unphysical case used earlier, where star formation in clusters is equivalent to the field ($f_\odot$=0, $\varepsilon=1$), the mock population synthesis estimates that approximately $1\pm1$ cluster associations out of 26 in our sample should be contributed by FRBs emitted from within galaxy clusters ($\gamma=0.0$, $N_\text{member}=1.0\pm0.9$, $\gamma=-0.5$, $N_\text{member}=0.8\pm0.8$, $\gamma=-1.0$, $N_\text{member}=0.9\pm1.0$). This number reflects only the cluster FRBs which are positively detected as associations by our method. The total number of cluster-emitted FRBs within the mock catalogs, including those that are undetected by our method is far higher, around $21\pm6$ ($\gamma=0.0$, $N_\text{complete}=20\pm5$, $\gamma=-0.5$, $N_\text{complete}=21\pm5$, $\gamma=-1.0$, $N_\text{complete}=22\pm7$). This poor completeness for member contributions is a natural consequence of inferring the probability a burst is foreground to a cluster from its DM (i.e., weighting the likelihood with $P(z<z_{\text{clust}}|\text{DM})$). Bursts emitted from within clusters have the lowest fractional excess DMs over the expectation at the cluster redshift, leaving them with the highest probabilities of being foreground to the cluster\footnote{Coincidentally, this means the number of member galaxy bursts persisting into the high purity sample in Fig. \ref{fig:tempProfile} is also low, vindicating the assumptions made in \S \ref{subsec:DMprofile}}. Furthermore, the high concentration we select for our score metric $\Sigma_s$ is more sharply peaked than the distribution of cluster members is, resulting in lower scores for cluster emitted FRBs in comparison to lensed FRBs. This unavoidably biases our search away from detecting FRBs within clusters, and so we expect the number of directly emitted FRBs in our sample to be small. It is possible to increase the completeness of the search to cluster member FRBs by using a lower concentration $\Sigma_s$ that more closely traces the distribution of cluster members \citep{budzynskiRadialDist2012}, however, this can also introduce a larger number of interlopers that may taint the purity of the sample overall, which is why we opt for the more concentrated $\Sigma_s$ map in this study.

\begin{figure}
    \centering
    \includegraphics[width=0.8\linewidth]{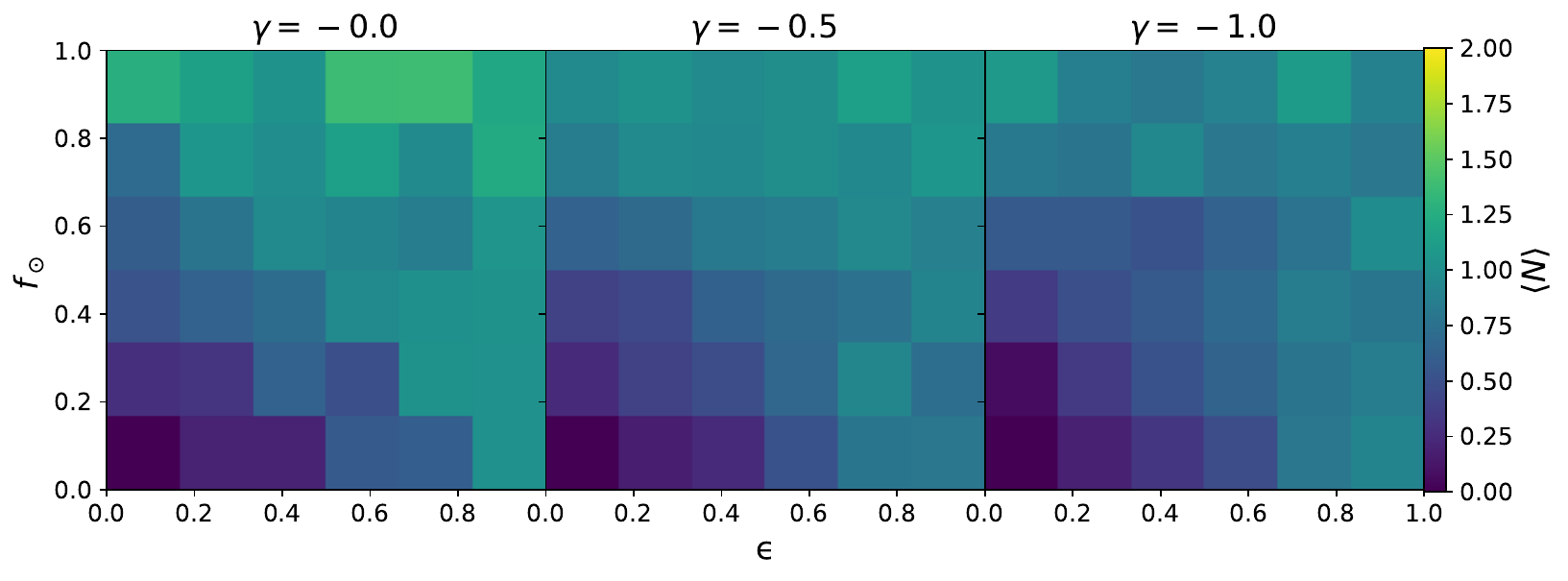}
    \caption{Expected number of cluster associations ($\langle N\rangle$) in our search contributed by FRBs from galaxy cluster member galaxies as a function of $f_\odot$ (the fraction of FRBs tracing stellar mass), and $\epsilon$ (the relative efficiency of star formation in galaxy clusters compared to the field), for three choices of energy function index ($\gamma$).}
    \label{fig:countsDirect}
\end{figure}

While the number of direct associations is small, it may provide a way to constrain the fraction of FRBs in alternative progenitor channels, due to the likelihood of suppressed star formation in clusters \citep{giodiniStellarTotalBaryon2009}. Fig. \ref{fig:countsDirect} demonstrates this, showing the expected number of associations contributed directly by galaxy cluster members as a function of both $f_\odot$ and $\epsilon$. Fig. \ref{fig:countsDirect} shows, as expected, that in the canonical regime where both $f_\odot$ and $\epsilon$ are small, that the contribution from galaxy cluster members is lowest, due to the suppression of the dominant FRB formation channel in clusters. As either $f_\odot$ or $\epsilon$ increases, the number of direct cluster associations also increases. Therefore, under the assumption that star formation is suppressed in clusters (i.e., $\epsilon$ is low), the number of FRBs associated with member galaxies provides a way to constrain $f_\odot$. Additionally, like the interlopers, the number of cluster FRBs shows only a weak dependence on $\gamma$, with steeper $\gamma$ values yielding slightly fewer associations. The relative independence of the cluster rate from $\gamma$ may be a boon to future constraints on $f_\odot$, however at present the primary challenge lies in identifying which out of the 26 FRBs are truly member contributions. 

Overall, we find that regardless of the precise population chosen, the direct contributions from member galaxies likely represent $\lesssim5\%$ of the sample of cluster associations we identify. We also estimate that our completeness to these direct contributions is only $\approx5\%$, with $20\pm6$ events in our initial CHIME/FRB baseband sample originating from cluster member galaxies. This is approximately equal to the total number of unmagnified interlopers FRBs expected to pass within $r_{500}$ of a foreground cluster, $20\pm4$. From the equivalence of these categories, we conclude that in the CHIME/FRB baseband sample, an FRB aligned within $r_{500}$ of a cluster with $M_{500}\geq10^{14}\,M_\odot$, is approximately equally likely to be from within that cluster or background to it. This stands in contrast to expectations from \cite{connorDeepSynopticArray2023a}, who suggest that FRBs aligned within $r_{200}$ of a cluster at $z_\text{clust}<0.5$, will be more likely to come from within the galaxy cluster itself than background to it, due to the relative number of galaxies within the cluster. We suggest that the validity of this conclusion rests on the mass of the cluster in question, $M_{\text{clust}}$, with the ratio between cluster members within $r_{200}$, $N_\text{clust}$, and background galaxies, $N_\text{background}$, shrinking for lower mass clusters as $N_{\text{clust}}/N_{\text{background}}\propto M_{\text{clust}}$, for a cluster with uniform density. As a result, the conclusions of \cite{connorDeepSynopticArray2023a} made for massive clusters with $N_{\text{clust}}\sim 100-200$, will not hold for the smaller, more prevalent clusters associated with most of our sample. Furthermore, DSA-110 and CHIME/FRB have differing instrumental responses and survey sensitivities. This changes the z-DM distribution of their observed bursts, which can impact the relative likelihoods of direct and background cluster FRBs, potentially causing the discrepancy in the fraction of FRBs emitted from member galaxies between the \cite{connorDeepSynopticArray2023a} sample and the sample reported here. This highlights the need to consider both instrumental selection and luminosity functions when considering cluster FRB rates in future analysis.


\subsubsection{Lensing}
For the fiducial case of a non-evolving energy function, the mock population synthesis described in \S \ref{subsec:forecasting} indicates that gravitational lensing from massive galaxy clusters in the \cite{wenCatalog158Million2024} sample should contribute $1\pm1$ cluster associations to our sample of 26 ($\gamma=0.0$, $N_\text{lensing}=1.6\pm1.5$, $\gamma=-0.5$, $N_{\text{lensing}}=1.5\pm1.4$, $\gamma=-1.0$, $N_{\text{lensing}}=1.1\pm1.1$). In this case, the completeness is higher at $\approx20\%$ with ($\gamma=0.0$, $N_\text{complete}=7.9\pm4.6$, $\gamma=-0.5$, $N_\text{complete}=7.4\pm4.2$, $\gamma=-1.0$, $N_\text{complete}=6.4\pm3.1$). 

\begin{figure}
    \centering
    \includegraphics[width=0.8\linewidth]{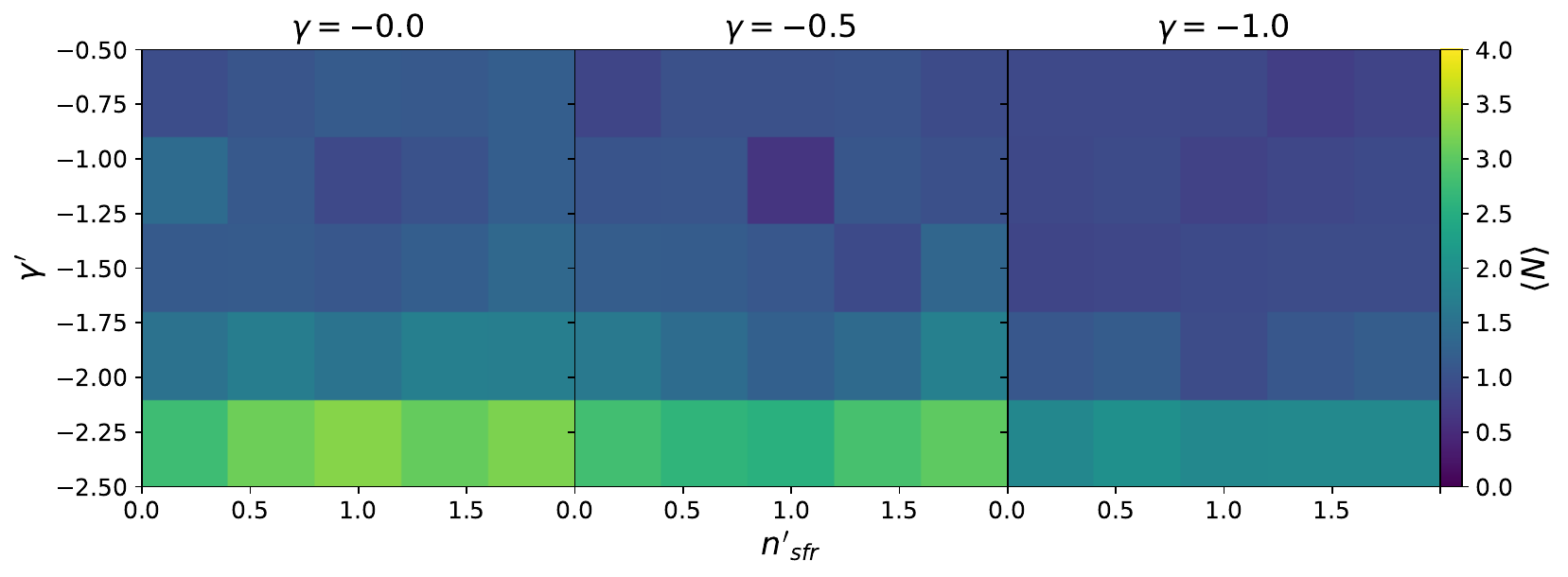}
    \caption{The expected number of cluster associations ($\langle N\rangle$) contributed by lensing from massive galaxy clusters ($M_{500}\gtrsim5\times10^14\,M_\odot$) for varying evolutions of the FRB population at redshifts $z\geq1$ from the baseline populations shown in Table \ref{tab:inputs} with $\gamma=0,\,-0.5, \,-1.0$ respectively.}
    \label{fig:evolution}
\end{figure}

Due to the magnification of distant sources, the expected contributions from gravitational lensing are potentially sensitive to high-redshift ($z\gtrsim 1$) evolution in the FRB population. To investigate this concept, we calculate the number of associations expected following our earlier method but for all combinations of $\gamma'$ and $n_{\text{SFR}}'$, as shown in Fig. \ref{fig:evolution}. As expected, the number of associations shows some dependence on the high-redshift evolution of the FRB population. In particular, changes to the slope of the energy function show the greatest effect, with steeper $\gamma$ values increasing the number of lensed FRBs by providing an abundance of lower energy bursts that can be magnified up to detectability. This indicates that, even in low numbers, observations of FRBs associated with cluster lenses can provide insight into the high-redshift behaviour of the FRB energy function. Conversely, the number of lensed FRBs shows little to no dependence on evolution in $n_{\text{sfr}}$, suggesting that future constraints on the redshift distribution via this method will be difficult. Overall, the contribution from cluster lensing to our detected sample, under any population model, is also subdominant, comprising only 5--10\% of the 26 burst sample detected here. 

In aggregate, the additional bursts from direct and lensing contributions increase the expected number of cluster associations to a total of $15.0\pm4.2$ for a non-evolving population tracing only star formation ($f_\odot=0$), with a cluster star-forming efficiency of $\epsilon=0.2$. While still below our true number of associations, this rate is within $3\sigma$, indicating that our observations do not require alternative progenitor channels or significant evolution in the FRB population. Our observations do, however, disfavour a number of population scenarios. Figures \ref{fig:sig0}, \ref{fig:sig05} and \ref{fig:sig10} in Appendix \ref{app:mocks} provide a complete distribution of the residuals between the trial populations and the observed sample of cluster associations. The key takeaway from these residuals is that across the $\{f_\odot, \epsilon, \gamma', n'_{\text{SFR}}\}$ parameter space, the differences between the observed and mock catalogs can be in excess of $8\sigma$, suggesting that cluster-associated FRBs can provide a powerful sample for constraining new complexities in the FRB population. For example, our cursory exploration suggests that a steep evolution in the energy index at high redshift, $\gamma'=-2.5$ can almost completely reconcile our simulations with the number of cluster associations we detect. Conversely, our forecasting favours a shallow base index $\gamma\geq-1.0$ overall, meaning we favour a higher average energy for FRBs, regardless of the broader population behaviour, but support the presence of numerous lower energy FRBs at higher redshifts. As we only perform 100 mock simulations in each case, the intrinsic noise in our exploratory mock catalog is large and therefore we do not attempt to derive formal constraints here. Our exploration demonstrates, however, that cluster FRBs have a strong ability to constrain complex population models. With the upcoming CHIME/FRB sensitivity upgrade and development of CHORD \citep{vanderlindeCHORD}, a constraining sample of several thousand FRBs localized to arcminute precision should be available in 2--4 years.

\begin{figure}
    \centering
    \includegraphics[width=\linewidth]{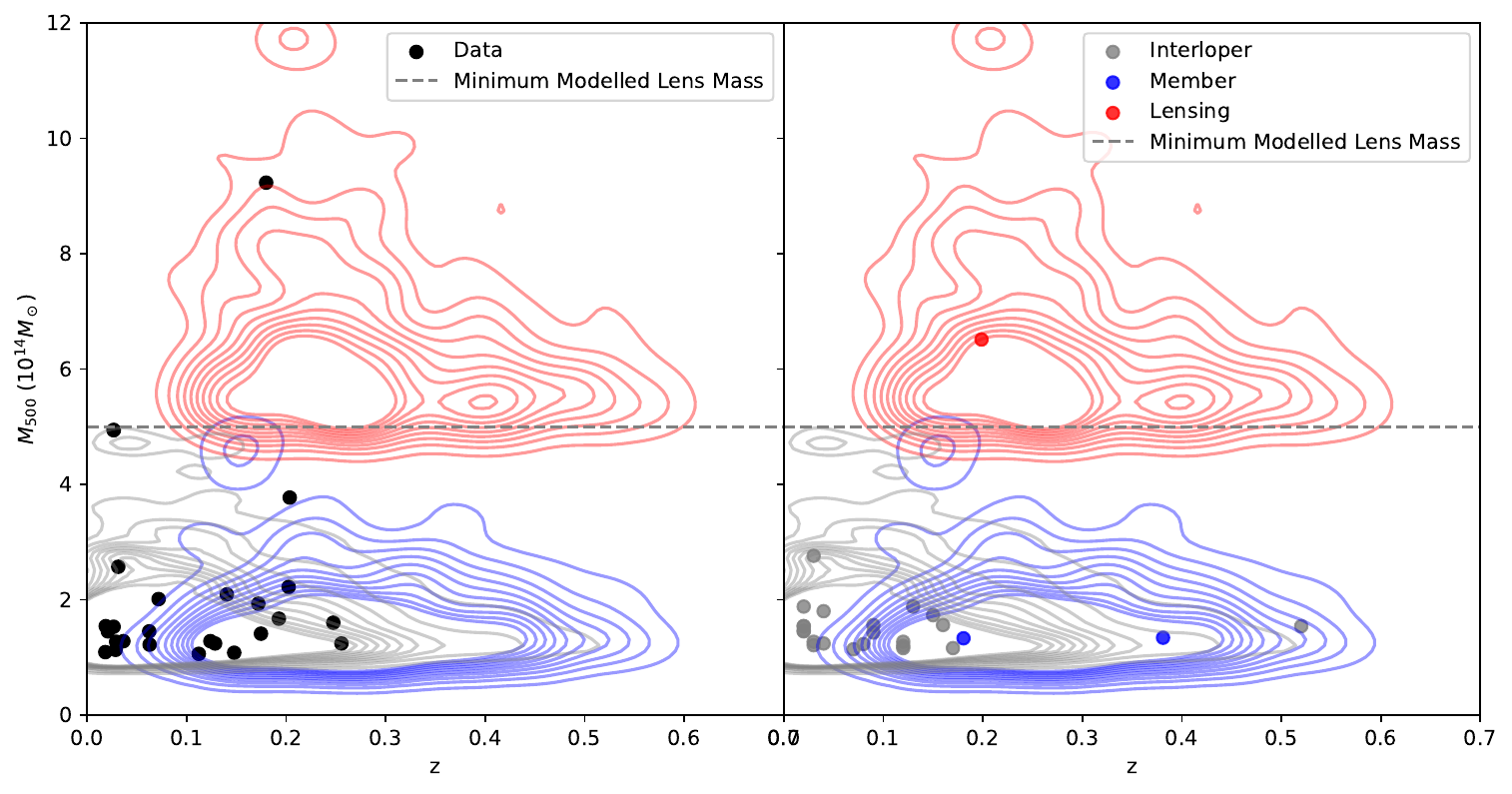}
    \caption{Properties of FRB associated clusters decomposed by source. Red contours highlight contributions from gravitational lensing, blue contours highlight FRBs emitted from within a galaxy cluster, and grey represent those of unmagnified interlopers randomly intercepting foreground galaxy halos. Each contour is built from the mock catalogs marginalized over all trial populations with each level corresponding to 10$\%$ of the whole population. The dashed grey line corresponds to minimum lens mass, below there are limited available lens models. \textit{Left:} Distribution of true clusters corresponding to the 26 FRB associations made in this study. \textit{Right:} A single mock catalog, qualitatively consistent with the observed population (left), but showing the typical contributions expected from each category. }
    \label{fig:contributionClusterMorph}
\end{figure}

Finally, the mock catalogs also provide estimates for the expected distributions for the masses and redshifts of the clusters associated with our FRBs. marginalizing over all trial populations, Fig. \ref{fig:contributionClusterMorph} shows the characteristic distribution of $M_{500}$ and $z_{\text{clust}}$ for cluster-associated FRBs contributed by members, lensed sources and random interlopers as contoured regions, and depicts a single mock sample for comparison with our observed sample. Fig. \ref{fig:contributionClusterMorph} shows that both the mock and observed samples are distributed over a similar region of parameter space, with most of the mock sample contributed by random interlopers intersecting low mass foreground galaxy clusters at redshifts $z_{\text{clust}}<0.3$. As discussed above, FRB emission from the cluster members themselves contributes only a small number of bursts, with cluster masses and redshifts expected to overlap with the interloper distribution, suggesting that identifying FRBs emitted from within clusters will be difficult without precision localization to a cluster member galaxy and careful consideration of the host association probabilities in a cluster environment. Conversely, contributions from cluster lensing can cause association with much higher mass clusters than expected for either the interlopers or member FRBs. This distinction serves as a powerful way to estimate which FRBs in the cluster sample are likely magnified by gravitational lensing. In our data, FRB\,20211113A is associated with a much more massive cluster than any other FRB in our sample, making it unlikely to be contributed either by direct emission from a member galaxy or by chance intersection. Instead, it is possible that FRB\,20211113A has been magnified by gravitational lensing from Abell 2218. We note that its DM may not be high enough for it to be background to Abell 2218 if it transits a highly magnified path in the inner region of the cluster, but nonetheless it represents a potential candidate worthy of further consideration. Candidate lensed FRBs of this type represent a potentially powerful new type of probe for dark matter microlensing \citep{lewisGravitationalMicrolensingTime2020a} and high-redshift FRB populations, and should be followed up with specific observations to search for echoes induced by strong lensing multiple imaging. Such echoes can occur on a variety of timescales, from the month to year timescales of cluster lensing \citep{kelly2015Sci, Rodney2021NatAs, Pierel2024ApJ, Pascale2025ApJ} to the ms timescales of microlensing \citep{munozLensingFastRadio2016, sammonsFirst, calvinPrimordial}, and can be confirmed definitively by coherence of the electric field between bursts \citep{kaderSim2024, leungWave2025} or by a high degree of similarity in burst spectra.

\section{Conclusion}\label{sec:conclusions}

To date, FRB population modeling has been largely successful in unifying the observations from disparate FRB surveys into a cohesive depiction of FRB behaviour. To extend this success further, we propose that population studies of future FRB samples consider specific FRB subpopulations in isolation, as a way to investigate emerging discrepancies in simple population models and to probe blind-spots in the data for which FRB surveys are, on average, insensitive. Specifically, we show that FRBs emitted from within or behind massive galaxy clusters are sensitive to the high-redshift regime and to the way in which the FRB population traces stellar mass, providing an insightful window for future population studies. To enable these studies, we demonstrate that the second CHIME/FRB baseband catalog must contain FRBs piercing foreground galaxy clusters. We show this by extracting a relation between impact parameter and DM that matches the characteristic shape and temperature for an ICM with an NFW profile. Using a Monte Carlo resampling to characterize the likelihood of cluster association, we isolate a subsample of 26 FRBs from the CHIME/FRB catalog that are likely to be associated with massive galaxy clusters, including one repeating FRB, and two FRBs that pierce the Coma Cluster at different impact parameters. 

We compare our results with a sophisticated set of simulations predicting the expected number of galaxy cluster associations due to emissions from their member galaxies, random interceptions, and gravitational lensing of background sources. We find that random interceptions by unmagnified FRBs are the dominant source of these cluster associations, but are insufficient to completely explain the number of observed associations at the 3$\sigma$ level, constituting the detection of a $1.4\pm0.4\%$ increase in the total number of detected FRBs in the second CHIME/FRB baseband catalog, due to the presence of massive galaxy clusters. For reasonable assumptions for the FRB population, we suggest that both FRBs emitted from within, and lensed by galaxy clusters are required to reconcile our simulations with the observed sample. We suggest that lensed and member FRBs should each account for $5-10\%$ of cluster associations that we make, with direct cluster emission only dominating over CHIME background rates for massive nearby clusters. Despite the small fraction of associations we expect to be contributed by direct and lensed modes, we find that such contributions will eventually allow the number of associated cluster FRBs to constrain the high-redshift behaviour of, and alternative progenitor channels in the FRB population. Not only does this highlight the importance of cluster FRBs as probes of sub-dominant features in the FRB population, but more generally, it shows that through statistical methods, unlocalized bursts remain a useful data product for understanding the FRB phenomena.

Finally, by characterising the typical redshift and mass of the FRB-associated clusters, we find that high-mass cluster associations ($M\geq5\times10^{14}\,M_\odot$) are far more likely to be contributed by gravitational lensing than by direct cluster emission or chance interception. We identify one such burst in our sample, FRB\,20211113A, which is aligned with the strong lensing cluster Abell 2218 with $M_{500}=9.2\times10^{14}\,M_\odot$. From a single unlocalized burst, there is no way to definitively confirm magnification from gravitational lensing, however we highlight FRB\,20211113A as an interesting candidate for further consideration.

\section{Acknowledgments}

We acknowledge that CHIME is located on the traditional, ancestral, and unceded territory of the Syilx/Okanagan people. We are grateful to the staff of the Dominion Radio Astrophysical Observatory, which is operated by the National Research Council of Canada. CHIME operations are funded by a grant from the NSERC Alliance Program and by support from McGill University, University of British Columbia, and University of Toronto. CHIME was funded by a grant from the Canada Foundation for Innovation (CFI) 2012 Leading Edge Fund (Project 31170) and by contributions from the provinces of British Columbia, Québec and Ontario. The CHIME/FRB Project was funded by a grant from the CFI 2015 Innovation Fund (Project 33213) and by contributions from the provinces of British Columbia and Québec, and by the Dunlap Institute for Astronomy and Astrophysics at the University of Toronto. Additional support was provided by the Canadian Institute for Advanced Research (CIFAR), the Trottier Space Institute at McGill University, and the University of British Columbia. The CHIME/FRB baseband recording system is funded in part by a CFI John R. Evans Leaders Fund award to IHS.

The AstroFlash research group at McGill University, University of Amsterdam, ASTRON, and JIVE is supported by: a Canada Excellence Research Chair in Transient Astrophysics (CERC-2022-00009); an Advanced Grant from the European Research Council (ERC) under the European Union’s Horizon 2020 research and innovation programme (`EuroFlash’; Grant agreement No. 101098079); an NWO-Vici grant (`AstroFlash’; VI.C.192.045); an NSERC Discovery Grant (RGPIN-2025-06681); an ERC Starting Grant (`EnviroFlash’; Grant agreement No. 101223057); and an NWO-Veni grant (VI.Veni.222.295).

M.W.S is a Fonds de Recherche du Quebec - Nature et Technologies (FRQNT) postdoctoral fellow and acknowledges support from the Trottier Space Institute Fellowship program. M.W.S also acknowledges support from the French government under the France 2030 investment plan, as part of the Initiative d’Excellence d’Aix-Marseille Université - AMIDEX (AMX-23-CEI- 088).
A.M.C. is a Banting Postdoctoral Fellow
K.W.M. is supported by NSF Grant No. 2510771 and receives lumbar support from the Adam J. Burgasser Chair in Astrophysics.
A.A acknowledges the support of the Natural Sciences and Engineering Research Council of Canada (NSERC) Undergraduate Summer Research Award (USRA).
A.P. is a Trottier Space Institute Postdoctoral Fellow.
V.M.K. holds the Lorne Trottier Chair in Astrophysics \& Cosmology, a Distinguished James McGill Professorship, and receives support from an NSERC Discovery grant (RGPIN 228738-13).
A.Pd. acknowledges support from the French government under the France 2030 investment plan, as part of the Initiative d'Excellence d'Aix-Marseille Universit\'e -- A*MIDEX (AMX-23-CEI-088).
K.T.M is supported by a FRQNT Master’s Research Scholarship.
A.P.C. is a Canadian SKA Scientist and is funded by the Government of Canada / est financé par le gouvernement du Canada.
P.S. acknowledges the support of an NSERC Discovery Grant (RGPIN-2024-06266).
D.M. acknowledges support from the French government under the France 2030 investment plan, as part of the Initiative d'Excellence d'Aix-Marseille Universit\'e -- A*MIDEX (AMX-23-CEI-088).
S.P.E.T is a Fonds de Recherche du Quebec - Nature et Technologies (FRQNT) doctoral fellow
A.B.P. acknowledges support by NASA through the NASA Hubble Fellowship grant \text{HST-HF2-51584.001-A} awarded by the Space Telescope Science Institute, which is operated by the Association of Universities for Research in Astronomy, Inc., under NASA contract \text{NAS5-26555}. A.B.P. also acknowledges prior support from a Banting Fellowship, a McGill Space Institute~(MSI) Fellowship, and a Fonds de Recherche du Quebec -- Nature et Technologies~(FRQNT) Postdoctoral Fellowship.
A.K. acknowledges the support of a grant (`2002729' or \url{https://doi.org/10.69777/2002729}) from the Fonds de recherche du Québec.
\bibliography{references}
\bibliographystyle{aasjournalv7}

\appendix
\section{Cluster Overlaps}
Table \ref{tab:Cluster} details the full list of cluster associations discovered in this work with properties from \cite{wenCatalog158Million2024}. The corresponding FRB properties can be found in Table \ref{tab:FRBs}. The DM$_{\text{ex}}$ column refers to the observed FRB DM with the Milky Way component predicted by NE2025 subtracted \citep{ockerNE2025UpdatedElectron2026}. Values of `---` represent quantities that could not be measured for various reasons listed by \cite{collaborationSecondCHIMEFRB2026a}.
\begin{table*}[h!]
\centering
\caption{Cluster Properties derived by \citet{wenCatalog158Million2024}}
\begin{tabular}{lccccccc}
\hline
FRB & Name & RA (J2000 deg) & Dec (J2000 deg) & $r_{500}$ (Mpc) & $M_{500}\,(10^{14}\,M_\odot)$ & $z$ & $N_{\mathrm{gal}}$ \\
\hline
FRB\,20190111B & J171907.2+132927 & 259.780 & 13.491 & 0.829 & 1.84 & 0.1510 & 15 \\
FRB\,20190111B & J171958.8+133058 & 259.995 & 13.516 & 0.723 & 1.08 & 0.1477 & 11 \\
FRB\,20190111B & J172011.4+133849 & 260.048 & 13.647 & 0.801 & 1.64 & 0.3374 & 10 \\
FRB\,20190701D & J072502.0+665632 & 111.259 & 66.942 & 0.809 & 1.64 & 0.0890 & 27 \\
FRB\,20190701D & J072658.7+664642 & 111.745 & 66.778 & 0.667 & 1.24 & 0.2555 & 12 \\
FRB\,20191019B & J160304.0+252540 & 240.767 & 25.428 & 0.753 & 1.41 & 0.1746 & 28 \\
FRB\,20191019B & J160319.8+252713 & 240.833 & 25.454 & 0.867 & 1.67 & 0.0895 & 28 \\
FRB\,20191030D & J071233.8+685330 & 108.141 & 68.892 & 0.747 & 1.28 & 0.1238 & 12 \\
FRB\,20191103A & J171056.3+394131 & 257.735 & 39.692 & 0.813 & 1.22 & 0.0627 & 35 \\
FRB\,20191105A & J080145.9+563312 & 120.441 & 56.553 & 0.814 & 1.27 & 0.0291 & 21 \\
FRB\,20191106A & J171120.3+431354 & 257.835 & 43.232 & 0.745 & 1.03 & 0.1719 & 20 \\
FRB\,20191106A & J171158.2+430646 & 257.992 & 43.113 & 0.839 & 1.93 & 0.1719 & 36 \\
FRB\,20191213A & J162838.3+393305 & 247.160 & 39.551 & 0.999 & 2.57 & 0.0312 & 60 \\
FRB\,Unknown & J085447.3+415458 & 133.697 & 41.916 & 0.896 & 2.09 & 0.1402 & 43 \\
FRB\,20201023D & J012339.9+331522 & 20.916 & 33.256 & 0.705 & 1.09 & 0.0182 & 13 \\
FRB\,20201125B & J010806.9+182752 & 17.029 & 18.464 & 0.804 & 1.67 & 0.1926 & 29 \\
FRB\,20201128B & J172714.5+521045 & 261.810 & 52.179 & 0.684 & 1.24 & 0.1285 & 18 \\
FRB\,20201128B & J172724.2+520215 & 261.851 & 52.038 & 0.705 & 1.08 & 0.1866 & 15 \\
FRB\,20210115A & J145357.5+033239 & 223.490 & 3.544 & 0.746 & 1.13 & 0.0285 & 7 \\
FRB\,20210130F & J133608.3+591223 & 204.035 & 59.206 & 0.893 & 2.01 & 0.0717 & 52 \\
FRB\,20210205B & J125612.2+274444 & 194.051 & 27.745 & 0.792 & 1.21 & 0.0231 & 30 \\
FRB\,20210205B & J130008.1+275837 & 195.034 & 27.977 & 1.262 & 4.94 & 0.0266 & 95 \\
FRB\,20210225A & J180633.1+513656 & 271.638 & 51.615 & 0.863 & 2.22 & 0.2024 & 34 \\
FRB\,20210411D & J125612.2+274444 & 194.051 & 27.745 & 0.792 & 1.21 & 0.0231 & 30 \\
FRB\,20210411D & J130008.1+275837 & 195.034 & 27.977 & 1.262 & 4.94 & 0.0266 & 95 \\
FRB\,20210706C & J125144.9+413759 & 192.937 & 41.633 & 0.769 & 1.60 & 0.2473 & 18 \\
FRB\,20211113A & J163549.3+661244 & 248.955 & 66.212 & 1.414 & 9.23 & 0.1798 & 173 \\
FRB\,20211117C & J010725.0+322445 & 16.854 & 32.413 & 0.813 & 1.54 & 0.0186 & 23 \\
FRB\,20211206C & J022745.9+281233 & 36.941 & 28.209 & 0.774 & 1.28 & 0.0363 & 29 \\
FRB\,20220328B & J224954.6+113631 & 342.478 & 11.609 & 0.837 & 1.53 & 0.0265 & 16 \\
FRB\,20220707D & J100840.4+135750 & 152.168 & 13.964 & 1.044 & 3.77 & 0.2035 & 55 \\
FRB\,20220906D & J135238.5+462059 & 208.161 & 46.350 & 0.803 & 1.45 & 0.0623 & 21 \\
FRB\,20221210H & J092539.1+362706 & 141.413 & 36.452 & 0.699 & 1.06 & 0.1121 & 14 \\
FRB\,20230720A & J002110.8+222126 & 5.295 & 22.357 & 0.794 & 1.45 & 0.0208 & 21 \\
\hline
\end{tabular}
\label{tab:Cluster}
\end{table*}

\begin{table*}[h!]
\centering
\caption{FRB Properties}
\begin{tabular}{lcccccc}
\hline
FRB & RA (J2000 deg) & Dec (J2000 deg) & DM (pc cm$^{-3}$) & DM$_{\rm ex}$ (pc cm$^{-3}$) & $\tau_{\rm 400\,MHz}$ (ms) & $F_\nu$ (Jy\,ms) \\
\hline
FRB\,20190111B & $259.993\pm0.004$ & $13.543\pm0.005$ & $1336.9$ & $1273.0$ & $0$ & $27.4\pm2.8$ \\
FRB\,20190701D & $111.766\pm0.005$ & $66.767\pm0.005$ & $933.3$ & $884.7$ & $11.16\pm0.54$ & $19.8\pm2.0$ \\
FRB\,20191019B & $240.76\pm0.02$ & $25.44\pm0.03$ & $1668.0$ & $1636.3$ & $14.49\pm2.22$ & $7.3\pm0.9$ \\
FRB\,20191030D & $108.145\pm0.005$ & $68.882\pm0.005$ & $248.7$ & $198.2$ & $3.43\pm0.25$ & $20.5\pm2.1$ \\
FRB\,20191103A & $257.709\pm0.007$ & $39.567\pm0.008$ & $1342.5$ & $1298.4$ & $0$ & $9.1\pm1.0$ \\
FRB\,20191105A & $120.799\pm0.004$ & $56.995\pm0.003$ & $721.3$ & $676.9$ & $0$ & $30.0\pm3.0$ \\
FRB\,20191106A & $257.952\pm0.006$ & $43.073\pm0.007$ & $1236.3$ & $1192.2$ & $3.11\pm0.16$ & $6.7\pm0.7$ \\
FRB\,20191213A & $247.143\pm0.003$ & $39.271\pm0.003$ & $978.5$ & $943.1$ & $3.64\pm0.13$ & $25.5\pm2.6$ \\
FRB\,20191220C & $133.750\pm003$ & $41.886\pm0.003$ & $840.9$ & $803.1$ &  $14.5\pm0.3$ & $59.0\pm6.0$ \\
FRB\,20201023D & $19.356\pm0.004$ & $32.836\pm0.004$ & $837.4$ & $803.0$ & $1.72\pm0.04$ & $30.3\pm3.1$ \\
FRB\,20201125B & $17.048\pm0.006$ & $18.474\pm0.007$ & $413.7$ & $378.4$ & --- & $11.8\pm1.3$ \\
FRB\,20201128B & $261.812\pm0.004$ & $52.165\pm0.003$ & $298.1$ & $253.0$ & $0$ & $34.9\pm3.5$ \\
FRB\,20210115A & $224.282\pm0.005$ & $4.017\pm0.005$ & $778.0$ & $747.4$ & $8.94\pm0.35$ & $30.7\pm3.4$ \\
FRB\,20210130F & $204.079\pm0.004$ & $59.245\pm0.004$ & $1574.2$ & $1544.6$ & $0$ & $13.4\pm1.4$ \\
FRB\,20210205B & $194.659\pm0.008$ & $27.604\pm0.009$ & $858.6$ & $836.9$ & $0$ & $15.5\pm2.0$ \\
FRB\,20210225A & $271.672\pm0.005$ & $51.632\pm0.004$ & $387.9$ & $334.5$ & $15.41\pm0.42$ & $35.3\pm3.6$ \\
FRB\,20210411D & $194.788\pm0.006$ & $27.123\pm0.006$ & $1224.6$ & $1202.4$ & $4.04\pm0.10$ & $20.3\pm2.1$ \\
FRB\,20210706C & $192.92\pm0.01$ & $41.62\pm0.01$ & $879.2$ & $856.5$ & $1.53\pm0.13$ & --- \\
FRB\,20211113A & $248.86\pm0.03$ & $66.25\pm0.02$ & $1467.9$ & $1428.2$ & $0$ & $5.8\pm0.9$ \\
FRB\,20211117C & $18.065\pm0.007$ & $33.744\pm0.007$ & $493.5$ & $445.5$ & $16.54\pm0.69$ & $27.5\pm2.8$ \\
FRB\,20211206C & $37.557\pm0.009$ & $28.54\pm0.01$ & $1018.8$ & $971.2$ & $0$ & $7.2\pm0.9$ \\
FRB\,20220328B & $341.834\pm0.009$ & $11.37\pm0.01$ & $1442.3$ & $1403.5$ & $6.45\pm0.45$ & $64.2\pm7.1$ \\
FRB\,20220707D & $152.15\pm0.02$ & $13.95\pm0.02$ & $551.7$ & $518.8$ & $24.83\pm2.63$ & $12.7\pm1.8$ \\
FRB\,20220906D & $208.424\pm0.006$ & $46.388\pm0.006$ & $526.8$ & $499.9$ & $0$ & $3.9\pm0.5$ \\
FRB\,20221210H & $141.30\pm0.01$ & $36.47\pm0.02$ & $1078.9$ & $1044.3$ & $4.88\pm0.55$ & $3.8\pm0.5$ \\
FRB\,20230720A & $4.453\pm0.005$ & $21.272\pm0.005$ & $532.6$ & $495.1$ & $32.48\pm0.43$ & $429.9\pm43.1$ \\
\hline
\end{tabular}
\label{tab:FRBs}
\end{table*}

\section{Monte Carlo Method}\label{app:montecarlo}
Our Monte Carlo method simulates samples of baseband FRB detections by CHIME/FRB that are statistically consistent with the spatial distribution of the sample used for this study and drawn from a range of underlying FRB populations that are consistent with current constraints. The default parameters of trial populations can be found in Table \ref{tab:inputs} with other tested populations described in \S \ref{sec:results}. In each case, the tested values are in reasonable agreement with population constraints from the first CHIME catalog \citep{shinInferringEnergyDistance2023} and therefore the DM distributions of the Monte Carlo sample will be in a reasonable agreement those of the first CHIME catalog (population results from the second CHIME catalog \citep{Jain2026} were unavailable at the time of this analysis).

\begin{figure}
    \centering
    \includegraphics[width=\linewidth]{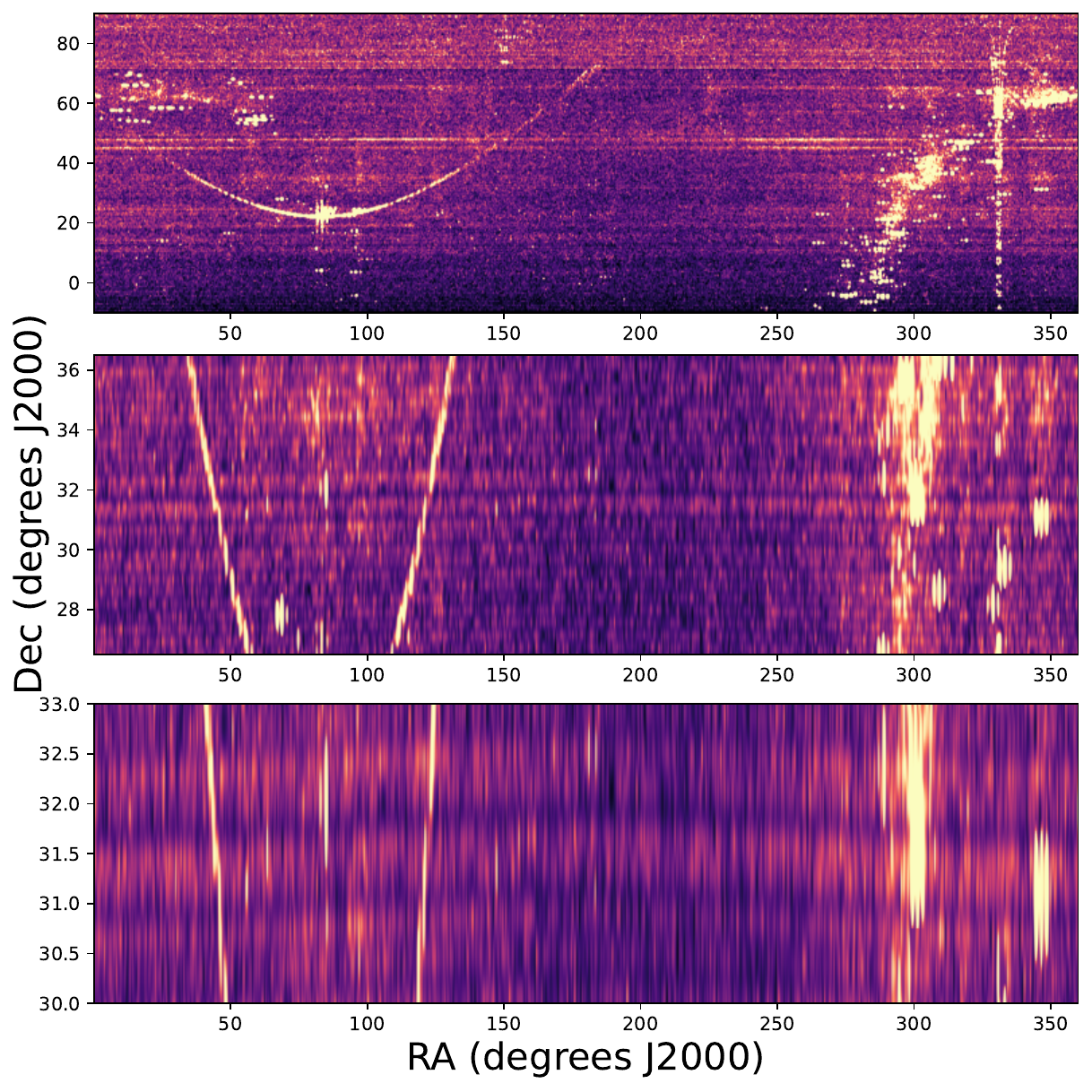}
    \caption{\textit{Top}: Spatial distribution of all CHIME/FRB detection candidates with S/N$\geq 7$ and DM$\geq50\,$pc cm$^{-3}$ observed by CHIME/FRB in 2022 and 2023, including FRBs, Galactic sources and terrestrial RFI. \textit{Middle, Bottom:} Declination zooms of the top plot.}
    \label{fig:CHIMEDetectionMap}
\end{figure}

To distribute our Monte Carlo sample consistently on the sky, we must first understand the spatial distribution of CHIME/FRB detections. To aid this discussion, we refer the reader to Fig. \ref{fig:CHIMEDetectionMap}, which plots the best estimates for the positions of all unfiltered candidate FRB events of S/N$\geq7$ and DM$\geq50\,$pc cm$^{-3}$, observed by CHIME/FRB in 2022 and 2023, convolved with a Gaussian representing the synthesized beam shape of CHIME/FRB. We select this period as it lies within the duration of the second baseband catalog, and we restrict ourselves to a year due to the volume of events. This is dominated by sub-threshold events ($\lesssim 8.5$) too faint to reliably categorize and includes many Galactic events from known pulsars and false positives from radio frequency interference (RFI) as well as all confirmed FRBs. As introduced above, the CHIME telescope has a cylindrical paraboloid reflector design giving it a narrow primary beam shape aligned along the meridian, extending over approximately $-11^\circ\leq\text{Dec}\leq110^\circ$ and peaking at zenith. This N-S beam sweeps the entire northern sky every day, with each position in the beam tracing all right ascensions at an approximately constant declination. This results in a distribution of detections that are approximately constant in right ascension but vary on large scales with declination. This declination dependence is both a function of the beam attenuation, with fewer astrophysical signals detected away from zenith, as can be seen in the low declinations in the top panel of Fig. \ref{fig:CHIMEDetectionMap}, and of exposure, with declinations above $\text{Dec}\gtrsim +70^\circ$ transiting twice every day due to the local elevation of the north celestial pole, as can be seen in the top panel of Fig. \ref{fig:CHIMEDetectionMap}. Additionally, the synthesized beams that are searched for FRBs are only sparsely distributed within the primary beam of CHIME, imprinting a small-scale declination dependence as well. This small-scale dependence is visualized in the bottom two panels of Fig. \ref{fig:CHIMEDetectionMap}, where the detections are slightly more concentrated in some rows. The separation of the beam centroids $(\sim0.5^\circ)$ is larger than the FWHM of a given synthesized beam ($\sim21.5'=0.36^\circ$) at the $600\,$MHz center frequency of the CHIME/FRB survey, leading to a dip in sensitivity between beams, particularly at higher frequencies, which introduces the small-scale declination dependence.

Considering both the large- and small-scale declination dependences for sensitivity and exposure described above, the distribution of probable positions for an FRB sample equivalent to our sample is a complex function of declination, but to first order, uniform in right ascension. Given the large number of FRB detections, we expect that this complex declination dependence will be well characterized by the true declination distribution of the detected bursts marginalized over RA. We may therefore draw a new distribution of positions for CHIME/FRB detections that is consistent with the survey's complex sky sensitivity by resampling each right ascension in the detected catalog from a uniform distribution and keeping each declination constant to within the localization uncertainty region of each burst.  

In detail, several other features in Fig. \ref{fig:CHIMEDetectionMap} are important to consider in order to resample the catalogs positions consistently with CHIME/FRB's detection capability. The most obvious of these features is the Galactic plane, which can be seen at both high and low right ascensions. In this context, the plane is represented by both the excess of noise signals introduced by the increased sky temperature of the Galaxy \citep{collaborationSecondCHIMEFRB2026a} and by the pulsed emission from Galactic pulsars. Additional noise from the plane reduces CHIME's overall sensitivity, limiting FRB detections at right ascensions on the Galactic plane, whereas, pulsar emission can act as a source of false positive FRB signals. As we will only be considering FRB detections from regions outside Planck's Galactic plane mask in our analysis, these effects on detectability can be disregarded. 

Outside the Galactic plane, some pulsars can still contribute candidate FRB signals. The most prominent example of this in Fig. \ref{fig:CHIMEDetectionMap} is the Crab pulsar, which is bright enough to be seen well into the side-lobes of CHIME's primary beam. The substantial offset of all synthesized beams from these side-lobes yields a spurious localization of the candidates, creating the apparent source arc centred around RA$=80^\circ$, $\text{Dec}\approx+20^\circ$. In all prominent cases of this, the pulsars are bright enough to be well known with well measured DMs, allowing all false-positive candidates along the arcs of known sources to be effectively filtered out. Furthermore, CHIME/FRB filters out a majority of potential Galactic radio transients by only considering signals with DMs larger than the total Milky Way contribution for that line of sight \citep{cordesNE2001NewModel2003, ywm16} to be potential FRBs. These filtering steps render any residual Galactic contribution to the candidate FRB rate of negligible impact to the uniformity of FRB detections to right ascension in regions outside the Planck Galactic plane mask.

Finally, perhaps the most subtle feature of Fig. \ref{fig:CHIMEDetectionMap} is the $\mathcal{O}(0.1)$ degree sinusoidal fluctuation in the beam declinations as a function of right ascension. This fluctuation is purely a coordinate system feature, caused by the drift in the Earth's rotation axis via precession in the $\approx22$ years between the mean observation time of the FRB catalog and the J2000 coordinate standard. The scale of the fluctuation is similar to the size of a synthesized beam in CHIME/FRB and is therefore important to consider when characterising the probable positions of detected FRBs. 

An aspect of the CHIME/FRB selection we have not addressed here is the correlation between declination and DM. As CHIME is more sensitive at declinations approaching zenith $\sim49^\circ$ the average DM at these declinations will be higher than those of lower elevation angles. modeling distinct populations for each declination throughout this work is computationally prohibitive and therefore we follow other CHIME studies \citep{shinInferringEnergyDistance2023,merryfieldInjectionSystemCHIME2023, McGregor2026} by marginalizing over this sky dependence of the DM distribution. As we generate independent Monte Carlo samples for each FRB's declination we expect this declination--DM covariance to be of marginal impact to our results. 

Considering all of the above, we use the following steps to simulate a catalog of FRB positions under the null hypothesis that all observed FRBs are foreground to the most distantly aligned galaxy cluster. We set the catalog size to 892 bursts, corresponding to the first detection of each unique FRB source in the second CHIME/FRB Baseband catalog within the region of the DECaLS map within $20\times r_{500}$ of an identified cluster from \cite{wenCatalog158Million2024}. The steps are as follows:
\begin{enumerate}
    \item Resample a FRB's RA from a uniform distribution between $0^\circ$ and $360^\circ$.
    \item Resample a FRB's Dec from a Normal distribution with a standard deviation equal to the baseband localization uncertainty for that burst.
    \item Offset each Dec according to the precession correction at that RA for the epoch of the FRB's observation.
    \item Draw a redshift and DM from the $p(z,\text{DM})$ distribution calculated using \texttt{z-DM} for the trial population.
    \item For samples in the masked region of the DECaLS map or with a redshift greater than that of a cluster within $3\times r_{500}$, repeat the above.
\end{enumerate}

Iterating this protocol, we build a Monte Carlo sample of $10^5$ FRB catalogs per population, characterising the probable locations of baseband localized FRB detections by CHIME/FRB in the unmasked region of the DECaLS exposure map under the null hypothesis that all FRB-cluster coincidences are chance alignments of foreground FRB sources.

\section{Rate Estimates}\label{app:rates}
\begin{figure}
    \centering
    \includegraphics[width=\linewidth]{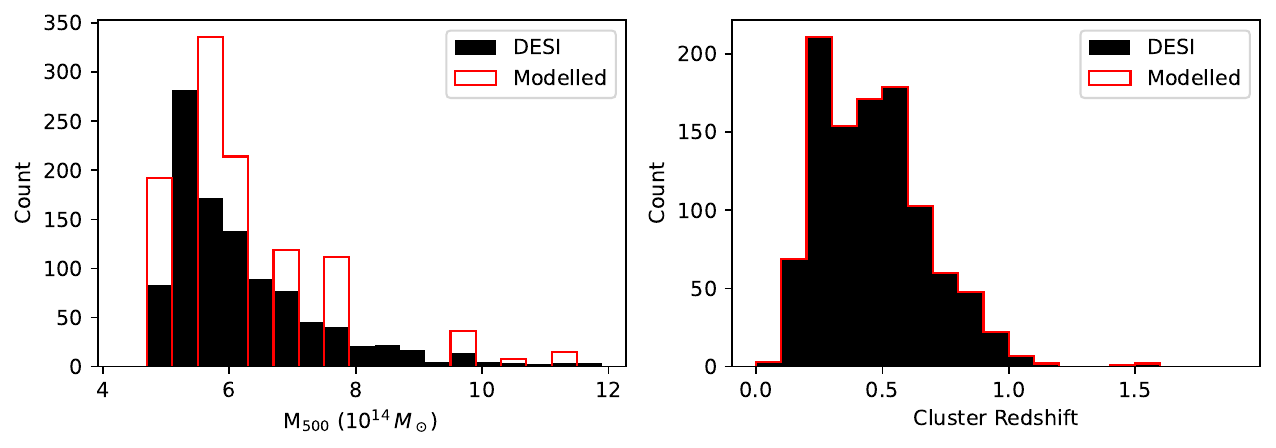}
    \caption{Galaxy cluster properties (\textit{left}: $M_{500}$, \textit{right}: redshift) from DECaLS compared to our representative model population in red.}
    \label{fig:modelledDemo}
\end{figure}

To represent the rate contributions from the most massive galaxy clusters in the \cite{wenCatalog158Million2024} sample, we must combine both the direct and lensing contributions. Using the calculations from \S \ref{subsec:forecasting}, the member contributions can be evaluated for each cluster using the mass and redshift measured by \cite{wenCatalog158Million2024}. For lensing, however, a majority of the \cite{wenCatalog158Million2024} clusters lack the requisite lensing model from which to calculate the lensing rate. We therefore estimate the lensing contributions by constructing a representative catalog using the 10 galaxy clusters models included by \cite{sammonsForecastingFastRadio2025}, but remapped to a range of redshifts $0.05\leq z\leq1.55$ at intervals of $\Delta z=0.1$. The coverage of this representative population is shown in Fig. \ref{fig:modelledDemo}. The figure shows perfect representation of the redshifts in the true cluster catalog, as the redshift of the galaxy cluster models can be controlled with arbitrary precision. Conversely, the masses of our modelled galaxy clusters are limited to those of the 10 high-mass clusters modelled by \cite{sammonsForecastingFastRadio2025}, which are in turn restricted by available lensing models and electron density profiles. As a result, our modelled population can only capture lensing effects from high-mass clusters and is concentrated in fewer masses overall. We expect that this distribution will still be a reasonable representation of the high-mass population, and as high-mass clusters make the most effective lenses, it provides useful insight into the importance of lensing in this context.

Using our modelled cluster sight lines in combination with the baseline unlensed rates, we can reconstruct our observed catalog of FRBs by drawing a representative sample from both cluster and non-cluster sightlines. To do so, we use the angular density of clusters in the centre of the DECaLS field (i.e., away from masked regions) to characterize how often various clusters are visible by CHIME, finding densities of $\approx 0.06$ per square degree for $M_{500}\geq5\times10^{14}\,M_\odot$, and $\approx 15$ per square degree for $M\geq10^{14}\,M_\odot$. Within the 1024, 21.5' FWHM synthesized beams which CHIME/FRB searches continually for FRBs, these densities correspond to, on average 6 beams containing clusters with $M_{500}\geq5\times10^{14}\,M_\odot$ and 1500 clusters satisfying $M_{500}\geq10^{14}\,M_\odot$ visible across all beams at any given moment. To sample our catalog, we then draw FRB detections as a function of time from Poisson distributions with mean rates given by the modelled detection rate in each beam. Specifically, at each time increment we uniformly sample six $M_{500}\geq5\times10^{14}\,M_\odot$ clusters from the modelled population seen in Fig. \ref{fig:modelledDemo}, and use their corresponding lensed rates for the six lensed beams. We then sample the remaining 1018 beams from the unlensed rate distribution. Finally, we draw 1500 random clusters uniformly from those with $M_{500}\geq10^{14}\,M_\odot$, and sample each for FRB detections to be added to the ensemble. This process is repeated until the total number of detected FRBs is equal to the sample that we use here (892 FRBs). Positive detections then have their redshifts and DMs sampled from the corresponding z-DM distribution, and their positions sampled either randomly following the Monte Carlo process outlined above or following the normalized NFW profile associated with the galaxy cluster that either emitted or magnified them. In the case of cluster emitted FRBs, the DMs are drawn from the unlensed z-DM distribution at the redshift of the cluster with an additional cluster DM added to them, drawn from either the cluster's corresponding electron density profile if it is available, or from a normal distribution $\mathcal{N}(200,100)$ pc\,cm$^{-3}$ with negative values floored to zero.

\section{Mock catalogs Compared to Observed Sample}\label{app:mocks}
Each of the below figures visualizes the significance of the residual between the total cluster associations predicted by the mock catalog and the observed sample, i.e $(\langle N_\text{mock}\rangle-26)/\sigma(N_\text{mock})$, where $\sigma(N_\text{mock})$ is the standard deviation in the number of associations for a given trial population. 

\begin{figure}
    \centering
    \includegraphics[width=0.8\linewidth]{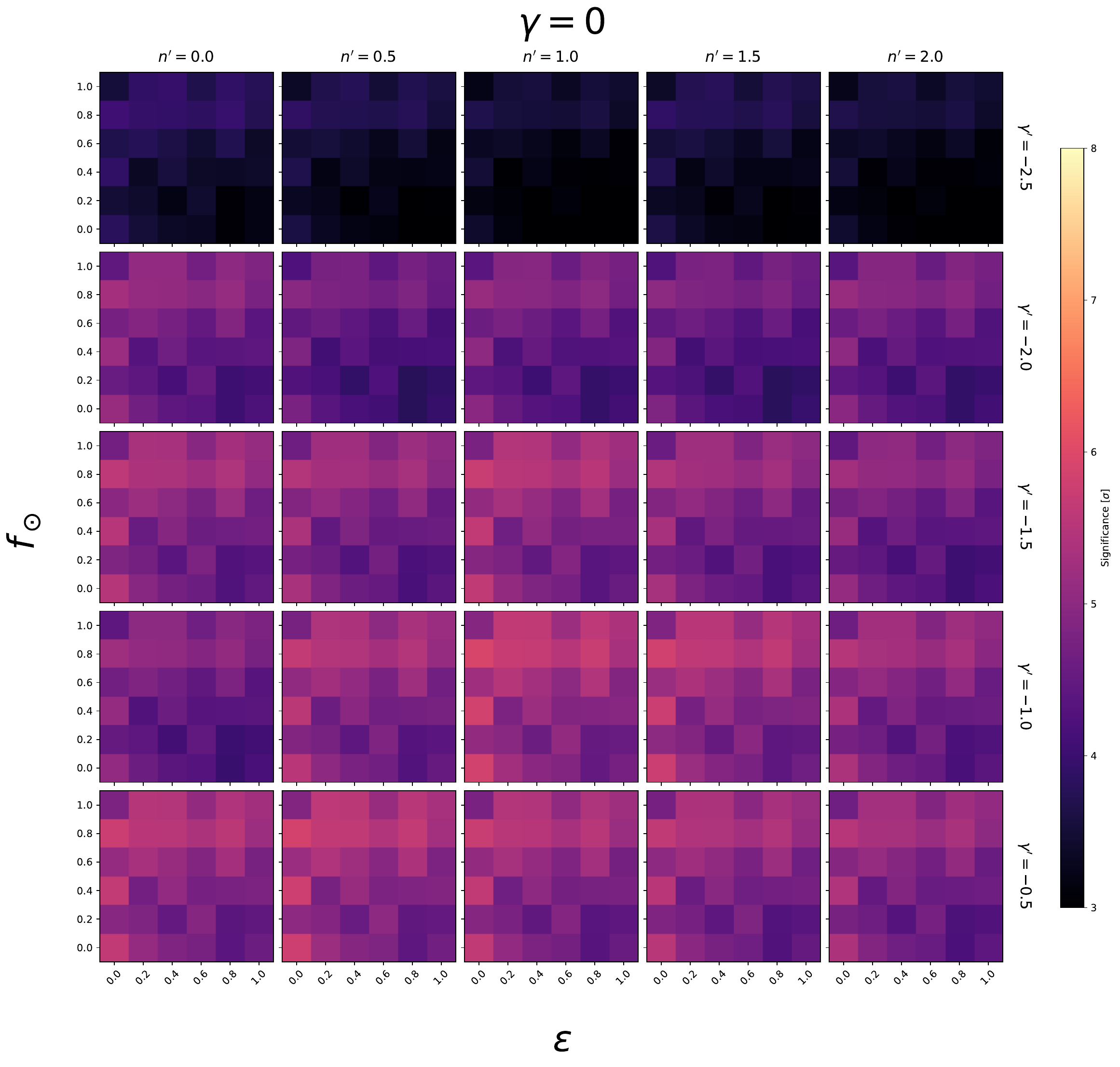}
    \caption{Significance of the residual between the number of cluster associations made in the observed sample and the mean of those in the mock catalogs for $\gamma=0$. The number of mock associations are calculated as the total number of associations from interlopers, member and lensing contributions for populations with a fraction of FRBs tracing stellar mass, $f_\odot$, relative galaxy cluster star formation efficiencies, $\varepsilon$, and high redshift evolutions in the energy and redshift distribution indices $\gamma'$ and $n'$ respectively.}
    \label{fig:sig0}
\end{figure}

\begin{figure}
    \centering
    \includegraphics[width=0.8\linewidth]{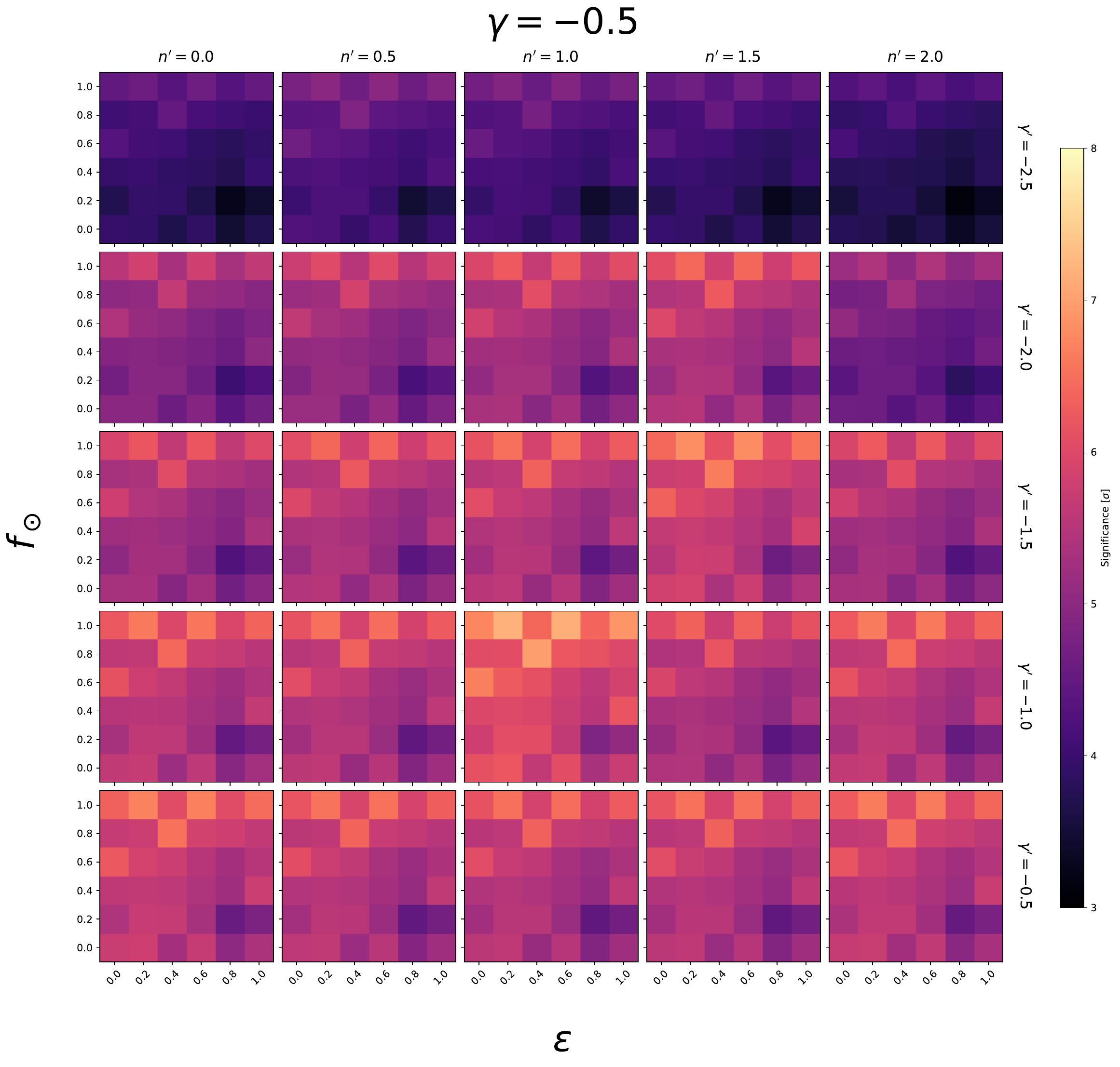}
    \caption{Significance of the residual between the observed sample and mock catalogs as in Fig. \ref{fig:sig0} for $\gamma=-0.5$.}
    \label{fig:sig05}
\end{figure}

\begin{figure}
    \centering
    \includegraphics[width=0.8\linewidth]{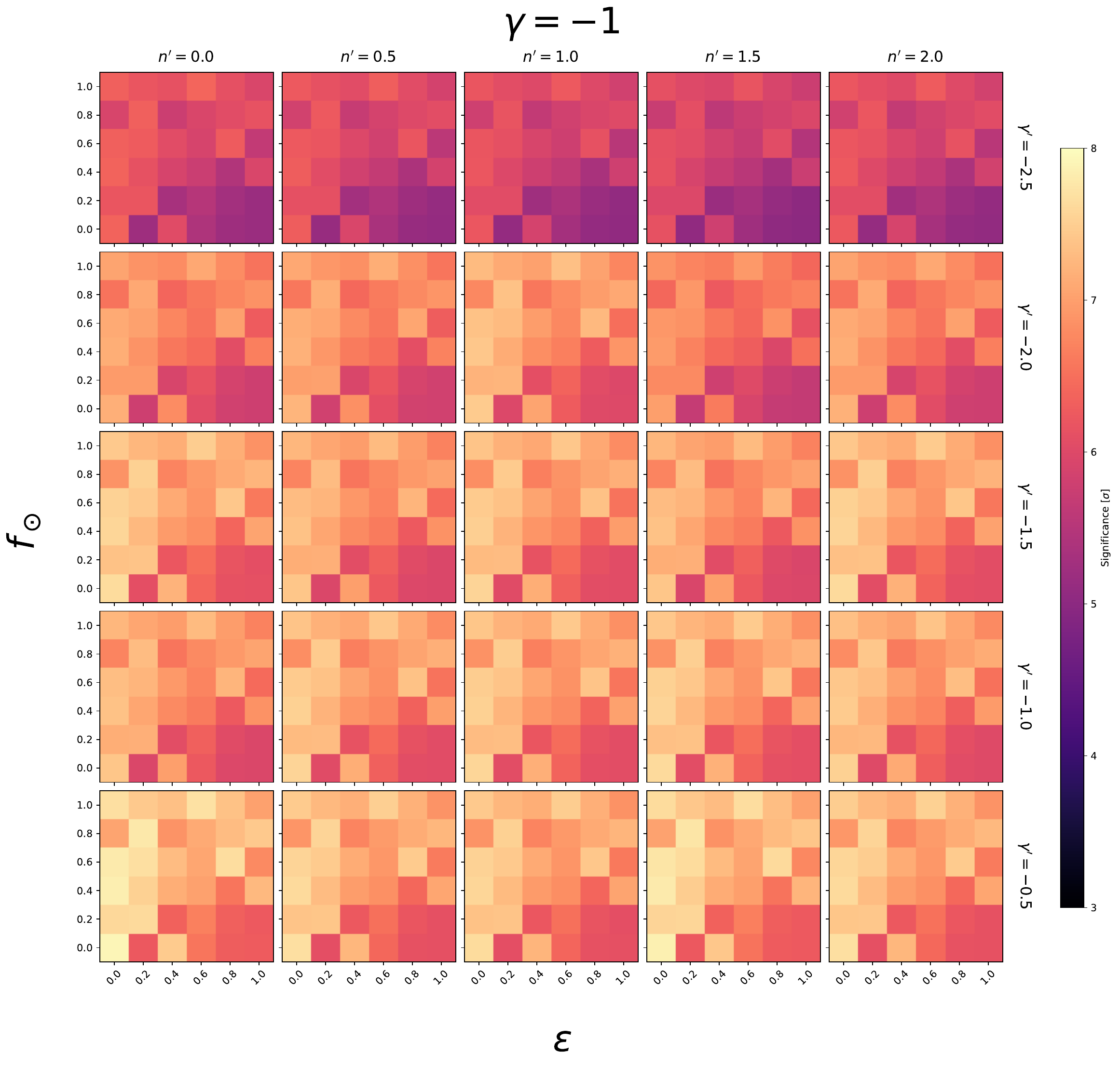}
    \caption{Significance of the residual between the observed sample and mock catalogs as in Fig. \ref{fig:sig0} for $\gamma=-1.0$.}
    \label{fig:sig10}
\end{figure}

\end{document}